\title{Time-varying neural network for stock return prediction}
\author{
        Steven Y. K. Wong\thanks{Steven Wong and Richard Xu are with School of Electrical and Data Engineering, University of Technology Sydney, Australia. Corresponding email: steven.ykwong87@gmail.com.},
        Jennifer S. K. Chan\thanks{Jennifer Chan and Lamiae Azizi are with School of Mathematics and Statistics, University of Sydney, Australia.},
        Lamiae Azizi\footnotemark[2], \\
        Richard Y. D. Xu\footnotemark[1]
}
\date{\today}

\documentclass[12pt]{article}

\usepackage[sort,round]{natbib} 
\usepackage{graphicx}
\usepackage{hyperref}
\usepackage{amsmath,amssymb,amsthm}
\usepackage{bm}
\usepackage{algorithm}
\usepackage{algpseudocode}
\usepackage{siunitx}
\usepackage{flafter}
\usepackage{booktabs}
\usepackage{textcomp}
\usepackage{color}
\DeclareMathOperator*{\argmin}{\arg\min}

\newcommand{\Ell}{\mathcal{L}}
\newcommand{\Tau}{\mathcal{T}}

\newcommand{\D}{\mathcal{D}}

\newcommand{\floor}[1]{\lfloor #1 \rfloor}
\newcommand{\norm}[1]{\left\lVert #1 \right\rVert}
\newcommand{\E}{\mathrm{E}}
\newcommand{\Var}{\mathrm{Var}}
\DeclareMathOperator{\R}{\mathbb{R}}
\DeclareMathOperator{\N}{\mathbb{N}}

\providecommand{\keywords}[1]
{
  \small	
  \textit{Keywords---} #1
}

\graphicspath{ {images/} }

\begin{document}
\maketitle

\begin{abstract}
	We consider the problem of neural network training in a time-varying context.
	Machine learning algorithms have excelled in problems that do not change over time. However, problems encountered in financial markets are often \emph{time-varying}.
	We propose the \emph{online early stopping} algorithm
	and show that a neural network trained using this algorithm can track a function changing with unknown dynamics.
	We compare the proposed algorithm to current approaches on predicting monthly U.S. stock returns and show its superiority.
	We also show that prominent factors (such as the size and momentum effects) and industry indicators, exhibit time varying predictive power on stock returns.
	We find that during market distress, industry indicators experience an increase in importance at the expense of firm level features.
	This indicates that industries play a role in explaining stock returns during periods of heightened risk.
\end{abstract}
\keywords{return prediction, deep learning, online learning, time-varying}

\section{Introduction} \label{sec:introduction}

The motivating application of this work is in predicting cross-sectional stock returns in a portfolio context.
At every interval, an investor forecasts expected return of assets and performs security selection.
A closely related problem is asset pricing --- a fundamental problem in financial theory\footnote{It is useful to remind readers that this paper is concerned with improving tools for stock return prediction, enabling practitioners to better select securities. Asset pricing, in an academic context, is more concerned with explaining the drivers of returns.}.
Asset pricing has been well studied.
A survey by \cite{Harvey:2016} documented over 300 cross-sectional factors published in journals.
However, literature has also documented evidence of time-variability of the true asset pricing model (also known as \emph{concept drift} in machine learning, \citealp{Gama:2014}).
\cite{PesaranTimmermann:1995} performed linear regressions with permutations of regressors on U.S. stocks and compared both statistical and financial measures for model selection.
Both predictability and regression coefficients of the selected model changed over time.
\cite{Bossaerts:1999} reported similar findings in international stocks.
So why do relationships change over time?
Changes in macroeconomic environment is one possibility.
Other explanations offered by \cite{AcademicDestroy:2016} relate to statistical bias (also called data mining bias in machine learning) and effects of arbitrage by investors (which the authors referred to as publication-informed trading).
Thus, it is unsatisfactory for a practitioner to learn a static model as out-of-sample performance can vary.

Recently, \emph{deep learning}\footnote{Deep learning is a subfield of \emph{machine learing}. An overview is provided in Section~\ref{sec:neural_networks}.} has made significant advances across a wide range of applications, such as achieving human-like accuracy in image recognition tasks \citep{FaceNet:2015} and translating texts \citep{Seq2Seq}.
By contrast, machine learning in financial markets is still in its infancy.
\cite{Weigand:2019} provided a recent survey of machine learning applied to empirical finance and noted that machine learning algorithms show promise in addressing shortcomings of conventional models (such as the inability to model non-linearity or handle large number of covariates).
Recent works have applied neural networks to the problem of cross-sectional stock return prediction (see, \citealp{Messmer:2017}; \citealp{Abe:2018}; \citealp{Gu:2020}).
\cite{Gu:2020} have modelled potential time-variability driven by macroeconomic conditions by interacting firm level features with macroeconomic indicators.
However, they do not consider all possible avenues of time-variability of asset pricing models, such as effects of trading as highlighted by \cite{AcademicDestroy:2016}.
For instance, \cite{ValueFailure:2019} noted that the prominent \emph{value} factor (\citealp{Rosenberg:1985}; \citealp{FamaFrench:1992}) has been unprofitable for almost 30 years --- a period that included multiple business cycles.
In fact, the authors noted that returns to the value factor have been negative since 2007, suggesting a change in the underlying relationship.

To address this, we propose the \emph{online early stopping} algorithm (henceforth, OES), for training neural networks that can adapt to a time-varying function.
Our problem is characterized by information release over time and iterative decision making.
Optimization in this context are called \emph{online} as decisions are made with the knowledge of past information but not the future.
In conventional neural network training, one of the hyperparameters is the number of optimization iterations $\tau$.
In OES, we propose to treat $\tau$ as a learnable parameter that varies over time ($t$), as $\tau_t$, and is recursively estimated over time.
We provide $\tau_t$ with a new meaning --- a regularization parameter that controls the amount of update neural network weights receive as new observations are revealed.
Thus, if consecutive cross-sectional observations are very different then we would expect $\tau_t$ to be small and the neural network is prevented from overfitting to any one period.
Conversely, a slowly changing function will have a high degree of continuity and we would expect the network to fit more tightly to each new observation.
Using this training algorithm, a neural network can adapt to changes in the data generation process over time.
For practitioners, we show that a neural network trained with OES can be a powerful prediction model and a useful tool for understanding the time-varying drivers of returns.

Neural network training is an optimization problem.
We draw on concepts in online optimization to provide a performance bound that is related to the variability of each period.
We do not assume any time-varying dynamics of the underlying function, a typical approach in online optimization.
The benefit of this approach is that it can track any source of variability in the underlying function, including macroeconomic, arbitrage-induced, market condition-induced or other unknown sources.
For instance, \cite{ValueFailure:2019} suggested that the negative return to the value factor was related to diminishing relevance of book equity as an accounting measure.
Such drivers would not have been captured by the macroeconomic approach in \cite{Gu:2020}.
Nonetheless, we acknowledge that a limitation of our approach is the inability to explain the source of variability.

We provide two evaluations of OES: 1) a simulation study based a data set simulated from a non-linear function evolving under a random-walk; 2) an empirical study of U.S. stock returns.
The empirical study is based on \cite{Gu:2020}, who compared several machine learning algorithms for predicting monthly returns of all U.S. stocks.
Majority of the data set were made available to the public and are used in this work.
We note that the setup in \cite{Gu:2020} is suboptimal for our portfolio selection problem for three reasons.
Firstly, (raw) monthly stock returns contain characteristics that complicate the forecasting problem, such as outliers, heavy tails and volatility clustering \citep{Cont:01}.
These characteristics are likely to impede a predictor's ability to learn.
Secondly, the data set in \cite{Gu:2020} contains stocks with very low market capitalization, are illiquid, and are unlikely to be accessible by institutional investors.
Thirdly, at the individual stock level, forecasting stocks' excess returns over risk free rate also encompasses forecasting market excess returns.
As practitioners are typically concerned with relative performance between stocks\footnote{In the simplest form, a long-only investor will hold a portfolio of the top ranked stocks and a long-short investor will buy top ranked stocks and sell short bottom ranked stocks. Thus, relative performance is relevant to practitioners.}, the market return component adds unnecessary noise to the problem of relative performance forecasting.
Thus, in addition to comparison with \cite{Gu:2020}, we also present results based on a more likely use case by practitioners, by excluding stocks with very low capitalization and forecasting cross-sectionally standardized excess returns.
We show that forecasting performance significantly improved based on this re-formulation.
We propose to measure performance using \emph{information coefficient} (henceforth, IC), a widely applied performance measure in investment management (\citealp{InfoCoeff:1974}; \citealp{GrinoldKahn:99}; \citealp{FabozziBook:2011:ch9}).
OES achieves IC of \SI{4.58}{\percent} on the U.S. equities data set, compared to \SI{3.82}{\percent} under an expanding window approach in \cite{Gu:2020}.

A summary of our contributions in this paper are as follows:
\begin{itemize}
	\item We propose the OES algorithm which allows a neural network to track a time-varying function.
		OES can be applied to an existing network architecture and requires significantly
		less time to train than the expanding window approach in \cite{Gu:2020}.
		In our tests, OES took 1/7 the time to train and predict as the expanding window approach of DNN.
		This has a practical implication as practitioners wishing to employ deep learning models have limited time between market close and next day's open to generate features and train new models, which is made worse if an ensemble is required.
	\item We show that firm features exhibit time-varying importance and that the model
		changes over time.
		We find that some prominent features, such as market capitalization (the size effect)
		display declining importance over time and is consistent with the findings of \cite{AcademicDestroy:2016}. This highlights the importance to have a time-varying model.
	\item We find that firm features, in aggregate, experience a fall in importance in predicting cross-sectional returns during market distress (e.g. Dot-com bubble in 2000-01).
	    Importance of sector dummy variables (e.g. technology and oil stocks) rose over the same period, suggesting importance of sectors is also time-varying. Our analysis indicates that sectors have an important role in predicting stock returns during market distress.
	    We expect this to be especially true if market stress impacts on certain sectors more than others, such as travel and leisure stocks during a pandemic.
	\item Using a subuniverse that is more accessible to institutional investors (by excluding microcap stocks), we show that OES exhibits superior predictive performance.
    	We find that mean correlation between predictions of OES and DNN is only \SI{35.9}{\percent} and monthly correlation is lowest immediately after a shock (e.g. recession).
    	We attribute this to OES adapting to the recovery which manifests as lower drawdown post Global Financial Crisis.
    \item We find that an ensemble formed by averaging the standardized predictions of the two models exhibits the highest IC, decile spread and Sharpe ratio.
    Thus, practitioners may choose to deploy both models in a complementary manner.
\end{itemize}

In the rest of this paper, we denote the algorithm of \cite{Gu:2020} as \emph{DNN} (Deep Neural Network) and our proposed Online Early Stopping as \emph{OES}. This paper is organized as follows.
Section~\ref{sec:prelim} defines our cross-disciplinary problem, and provides overviews of neural networks and online optimization.
Section~\ref{sec:online_deep_learn} outlines our main contribution of this paper --- the proposed OES algorithm
which introduces time-variations to the neural network.
Simulation results are presented in Section~\ref{sec:simulation}, which demonstrates the effectiveness of OES in tracking a time-varying function.
An empirical study on U.S. stock returns is outlined in Section~\ref{sec:us_equities}. Finally, Section~\ref{sec:conclusion} discusses the empirical finance problem and concludes the paper with some remarks.

\section{Preliminaries} \label{sec:prelim}

\subsection{Definitions} \label{sec:definitions}

Similar to a classical online learning setup, a player iteratively makes portfolio selection decisions at each time period.
We call this iterative process \emph{per interval training}.
There are $n$ stocks in the market, each with $m$ features, forming input matrix
$\bm{X}_t \in \R^{n{\times}m}$ at time $t = 1, ..., T$. The $i$-th row in $\bm{X}_t$ is feature vector $\bm{x}_{t,i}$ of stock $i$.
To simplify notations, we define return of stock $i$ as return over the next period, i.e.,
$r_{t,i} = (p_{t+1,i} + d_{t+1,i}) / p_{t,i} - 1$, where $p_{t,i}$ is price at time $t$ and $d_{t,i}$ is dividend at $t$ if a dividend is paid, and zero otherwise.
Player predicts stock returns $\hat{\bm{r}}_t \in \R^n$ by choosing $\bm{\theta}_t \in \Theta$, which parameterizes prediction function $F: \R^{n{\times}m} \mapsto \R^n$.
Market reveals $\bm{r}_t$ and, for regression purposes, investor incurs squared loss,
\[ J_t(\bm{\theta}_t) = \frac{1}{n} \sum_{i=1}^{n} (r_{t,i} - F_i)^2, \]
where $F_i$ is the $i$-th element of vector $F(\bm{X}_t; \bm{\theta}_t)$.
The true function $\phi_t: \R^{n \times m} \mapsto \R^n$ drifts over time and is approximated by $F$ with time-varying
$\bm{\theta}_t$.
Investor's objective is to minimize loss incurred by choosing the best $\bm{\theta}_t$ at time $t$ using observed history up to $t-1$.
Both the function form and time-varying dynamics of $\phi_t$ are not known. Hence a neural network is used to model the cross-sectional relationship at each $t$
and the time-variability is formulated as a network weights tracking problem.
The loss function $J$ verifies the same assumptions adopted in \cite{DTSSGD:2019}, which are:
\begin{itemize}
    \item $J_t$ is bounded: $|J_t| \le D; D > 0$,
    \item $J_t$ is L-Lipschitz: $|J_t(\bm{a}) - J_t(\bm{b})| \le L\norm{\bm{a} - \bm{b}}; L > 0$, 
    \item $J_t$ is $\beta$-smooth: $\norm{\nabla J_t(\bm{a}) - \nabla J_t(\bm{b})} \le \beta\norm{\bm{a} - \bm{b}}; \beta > 0$.
\end{itemize}
We denote the gradient of $J_t$ at $\bm{\theta}_t$ as
$\nabla J_t(\bm{\theta}_t)$ and stochastic gradient as
$\hat{\nabla}J_t(\bm{\theta}_t) = \E[\nabla J_t(\bm{\theta}_t)]$, or where the context
is obvious, $\nabla_t$ and $\hat{\nabla}_t$ respectively.

As performance measure, \cite{Gu:2020} used pooled $R_{oos}^2$ without mean adjustment in the denominator,
\[ R_{oos}^2 = 1 - \frac{\sum_{(t,i) \in \D_{oos}}(r_{t,i} - \hat{r}_{t,i})^2}{\sum_{(t,i) \in \D_{oos}} r_{t,i}^2}, \]
where $\D_{oos}$ is the pooled out-of-sample data set covering January 1987 to December 2016 in the empirical study.
There are several shortcomings with this performance measure.
The number of stocks in the U.S. equities data set starts from 1,060 in March 1957, peaks at over 9,100 in 1997, and falls to 5,708 at the end of 2016.
A pooled performance metric will place more weight on periods with a higher number of stocks.
An investor making iterative portfolio allocation decisions would be concerned with accuracy \emph{on average over time}.
Moreover, asset returns are known to exhibit non-Gaussian characteristics \citep{Cont:01}.
Summary statistics of monthly U.S. stock returns are provided in Table~\ref{tab:return_describe} (in Section~\ref{sec:us_equities}), which clearly confirms the existence of considerable skewness and time-varying variance.
Therefore, we provide three additional metrics.
The first metric is the \emph{information coefficient} (IC), defined as the cross-sectional Pearson's correlation\footnote{Rank IC, which uses Spearman's rank correlation instead of Pearson's, is also used in practice.} between stock returns and predictions.
The time series of correlations is then averaged to give the final score.
IC was first proposed by \cite{InfoCoeff:1974} and is widely applied in investment management for measuring predictive power of a forecaster or an investment strategy (\citealp{GrinoldKahn:99}; \citealp{FabozziBook:2011:ch9}).
The second metric is the annualized Sharpe ratio, calculated as,
\[ {SR} = \frac{12\times\E[P_{t\in\D_{oos}}]}{\sqrt{12\times\Var[P_{t\in\D_{oos}}]}}, \]
where $P_{t\in\D_{oos}}$ is the difference between average realized monthly returns of decile 10 and decile 1 at $t$, sorted on predicted returns.
The third metric is the average monthly $R^2$, where denominator is adjusted by the cross-sectional mean, as a conventional complement to $R_{oos}^2$.

\subsection{Feedforward neural networks} \label{sec:neural_networks}

An overview of neural networks is provided in this section. Interested readers are referred to \cite{Goodfellow:2016} for a comprehensive review.

\begin{figure}[h]
\caption{An illustration of a fully connected network with two hidden layers. $n_{\ell}$ refers to the
number of nodes in $\ell$-th layer. Arrows indicate direction of flow for the output value of the respective node.}
\label{fig:neural_network}
\includegraphics[width=12cm]{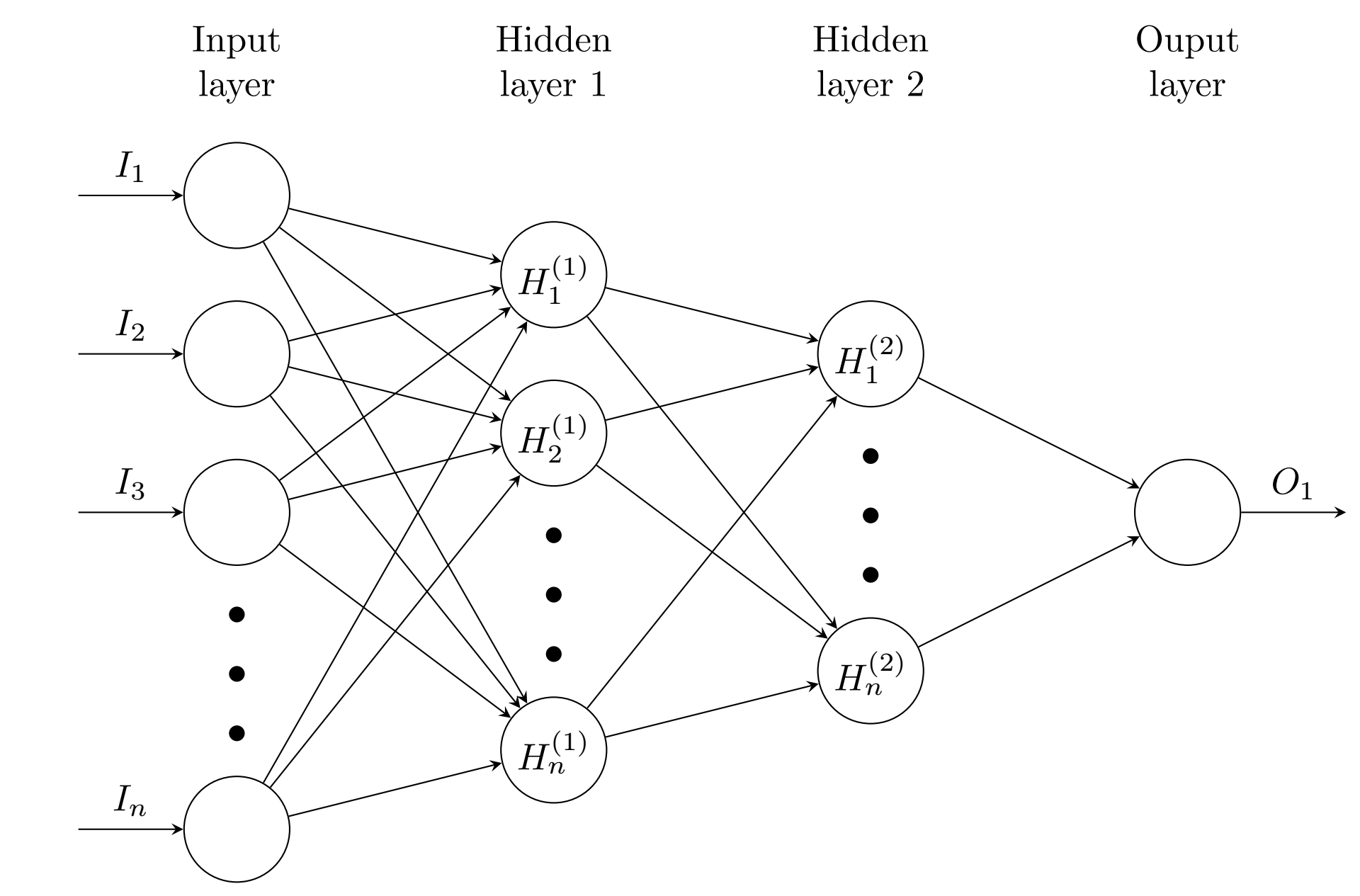}
\centering
\end{figure}

Neural networks are a broad class of high capacity models which were inspired by the biological brain and can theoretically learn any function (known as the \emph{Universal Approximation Theorem}, see \citealp{Hornik:1989}; \citealp{Cybenko:1989}; \citealp{Goodfellow:2016}).
A common form, the feedforward network, also known as \emph{multilayer perceptrons} (MLP),
is a subset of neural networks which forms a finite acyclic graph \citep{Goodfellow:2016}.
There are no loop connections and values are fed forward, from the input layer to hidden layers, and to the output layer.
The word `deep' is prefixed to the name (e.g. deep feedforward network
or deep neural network) to signify a network with many hidden layers, as illustrated in Figure~\ref{fig:neural_network}.
A feedforward network is also called a fully
connected network if every node has every node in the preceding layer connected to it.
The output of each layer acts as input to the next layer and loss is `backpropagated' by taking the partial derivative of loss with respect to weights at each layer.
Each layer consists of activation function $f$ (e.g. \emph{rectified linear unit}, defined as $f(x) = \max(x, 0)$),
weights $\bm{W}$, bias $b$, and output $f(\bm{x}; \bm{W}, b) = f(\bm{x}^T\bm{W} + b)$. The $\ell$-th layer of the network is denoted as $f^{(\ell)}$.
For brevity, we drop the layer designation,
and denote the entire network as $F$ and weight vector set $\bm{\theta} = \bigcup_{\ell=1}^{\Ell} \{\bm{W}^{(\ell)}, b^{(\ell)}\}$, where $\Ell$ is the number of layers. The network is trained with \emph{stochastic gradient descent} (or variants) at time $t$ (but dropping the subscript $t$ for simplicity as the context is clear),
\[ \bm{\theta}_{k+1} = \bm{\theta}_k - \eta \hat{\nabla}J(\bm{\theta}_k), \]
where $\bm{\theta}_k$ is the weight vector at optimization iteration $k$ (also called \emph{epochs}) and
$\eta$ is step size.

At time $t$, $\tau_t$ denotes the number of optimization iterations that are used to train the network and is found by monitoring loss on a validation set.
This procedure is called \emph{early stopping} (\citealp{Morgan:Bourland:1990}; \citealp{Reed:1993}; \citealp{Prechelt:1998}; \citealp{Mahsereci:2017:EarlyStop}).
Training is stopped when the validation loss decreases by less than a predefined amount, called \emph{tolerance}.
Early stopping can be seen as a regularization technique which limits the optimizer to search in the parameter space near the starting
parameters (\citealp{Sjoberg:Ljung:1995}; \citealp{Goodfellow:2016}).
In particular, given optimization steps $\tau$, the product $\eta \tau$ can be interpreted as the
effective capacity which bounds reachable parameter space from $\bm{\theta}_0$, thus behaving like $L_2$ regularization \citep{Goodfellow:2016}.

For time series problems where chronological ordering is important, popular approaches include expanding window (each new time slice is added to the panel data set)
and rolling window (the oldest time slice is removed as a new time slice is added, \citealp{Rossi:Inoue:2012}). Instead of randomly splitting training and test sets,
the \emph{out-of-sample} procedure\footnote{As described in \cite{Bergmeir:2018}.} can be used where the end of the series is withheld for evaluation.
This is unsatisfactory in the context of stock return prediction for two reasons. First, each time period is drawn from a different data distribution $\D$
(hereon denoted as $\D_t$ for data set drawn at time $t$, or $\D_{oos}$ for all periods in the out-of-sample data set). A pooled regression with window size $w$ effectively assumes data at
$t+1$ is drawn from the average of the past $w$ observations.
Secondly, if data is scarce in terms of time periods, estimates for optimal optimization steps $\hat{\tau}_t$ can have large stochastic error.
For instance, monthly data with a window size of 12 months and 3:1 training-validation split.
$\hat{\tau}$ is estimated using only 3 months of data.
To the best of our knowledge, there is no procedure for adapting early stopping in an online context with time-varying dynamics.

\subsection{Online optimization} \label{OnlineConvexOpt}

Optimizing network weights to track a function evolving under unknown dynamics is an online optimization problem.
A discussion on relevant concepts in online optimization is provided. Interested readers are encouraged to read \cite{Shalev-Shwartz:2012} for an introduction. In online optimization literature, iterate is often denoted as $x_t$ and loss function as $f_t$.
We have used $\bm{\theta}_t$ as iterate to be consistent with our parameter of interest and $J_t$ as loss function to avoid conflict with our use of $f$ as activation function.

Online optimization and its related topics have been well researched.
Applications of online optimization in finance first came in the form of the \emph{Universal Portfolios}  by \cite{UniversalPortfolios:1991}.
However, most of the early works in online optimization are on the convex case and assume each draw of loss function $J_t$ is from the same distribution (in other words, $J_t$ is stationary).
These assumptions are not consistent with our problem.
Recently, \cite{Hazan:2017} extended online convex optimization to the non-convex and stationary case.
This was further extended by \cite{DTSSGD:2019} to the non-convex and non-stationary\footnote{Non-stationarity in online optimization literature refers to time-variability of loss function $J_t$.} case, with the proposed \emph{Dynamic Exponentially Time-Smoothed Stochastic Gradient Descent} (DTS-SGD) algorithm.
Non-convex optimization is NP-Hard\footnote{In computer science, NP-Hard refers a class of problems where no known polynomial run-time algorithm exists.}.
Therefore, existing non-convex optimization algorithms focus on finding local minima \citep{Hazan:2017}.
For this reason, one difference between online \emph{convex} optimization and online \emph{non-convex} optimization is that the former focuses on minimizing sum of losses relative to a \emph{benchmark} (for instance, the minimizer over all time intervals $\bm{\theta}^* = \argmin_{\bm{\theta}\in\Theta} \sum_t J_t(\bm{\theta})$ is one of the most basic benchmarks), and the latter focuses on minimizing sum of gradients (e.g. $\sum_t \nabla J_t(\bm{\theta}_t)$).
This optimization objective is called \emph{regret}.
Readers familiar with time-series analysis might be taken aback by the lack of parameters in a typical online optimization algorithm.
This is due to the game theoretic approach of online optimization and the focus on worst case performance guarantees, as opposed to the average case performance in statistical learning.
Regret bounds are typically functions of properties of the loss function (e.g. convexity and smoothness) and are dependent on environmental assumptions.

At each interval $t$, DTS-SGD updates network weights using a time-weighted sum of past observed gradients. Time weighting is controlled by a forget factor $\alpha$.
In analyzing DTS-SGD, we note two potential weaknesses. Firstly, neural networks are notoriously difficult to train.
Geometry of the loss function is plagued by an abundance of local minima and saddle points (see Chapter~8.2 of \citealp{Goodfellow:2016}).
Momentum and learning rate decay strategies (for instance, \citealp{Sutskever:2013}; \citealp{ADAM:2015}) have been introduced which require multiple passes over training data, adjusting learning rate each time to better traverse the loss surface.
DTS-SGD is a single weight update at each time period which may have difficulties in traversing highly non-convex loss surfaces.
Secondly, during our simulation tests, we observed that loss can increase after a weight update.
One possibility is that a past gradient is taking the weights further away from the current local minima.
This is particularly problematic for our problem as stock returns are very noisy.

\section{Online early stopping} \label{sec:online_deep_learn}

\subsection{Tracking a restricted optimum} \label{sec:restricted_optimum}

We start by providing an informal discussion of the algorithm.
Neural networks are universal approximators.
That is, it can approximate any function up to an arbitrary accuracy.
Thus, given a network structure and a time-varying function, network weights trained with data from a single time interval (i.e., a cross-sectional slice of time) neatly summarizes the function at that interval and the Euclidean distance between consecutive sets of weights can be interpreted as the amount of variations in the latent function expressed in weight space.
Simply using $\bm{\theta}_{t-1}$ to predict on $t$ will lead to an overfitted result.
To illustrate, suppose $\bm{\theta}_t \in \R$, $\bm{\theta}_0 = 0$ and $\bm{\theta}_t$ alternates in a sequence of $\{1, -1, 1, -1, ... \}$.
Then, it is clear that using $\bm{\theta}_1 = 1$ to predict on $t = 2$ will lead to a worse outcome than using $\bm{\theta}_0 = 0$.
In this scenario, the optimal strategy is to never update weights (or scale updates by zero).
Generally, the optimal policy is to regularize updates such that the network is not overfitted to any single period.

In the rest of this section, we present our main theoretical results.
Formally, our goal is to track the unobserved minimizer of $J_t$, a proxy for the true asset pricing model, as closely as possible.
In regret analysis, it is desirable to have regret that scales sub-linearly to $T$, which leads to asymptotic convergence to the optimal solution.
\cite{Hazan:2017} demonstrated that in the non-convex case, a sequence of adversarially chosen loss functions can force any algorithm to suffer regret that scales with $T$ as $\Omega\left(\frac{T}{w^2}\right)$\footnote{In computer science, $\Omega$ notation refers to the lower bound complexity.}.
Locally smoothed gradients (over a rolling window of $w$ loss functions) were used to improve \emph{smoothed regret}, with a larger $w$ advocated by \cite{Hazan:2017}.
\cite{DTSSGD:2019} extended this to use rolling weighted average of past gradients which gives recent gradients a higher weight to track a dynamic function.
Inevitably, smoothing will track a time-varying minimizer with a tracking error that is proportionate to $w$ and the forget factor.

\begin{figure}[htbp]
\caption{At each optimization iteration, weights can be visualized as moving along the direction of $-\nabla J_{t-1}(\bm{\theta}')$. On the left, optimization should continue until $-\nabla J_{t}(\bm{\theta}')$ is perpendicular to $-\nabla J_{t-1}(\bm{\theta}')$. On the right, optimization should terminate.}
\label{fig:illustration_oes}
\includegraphics[width=12cm]{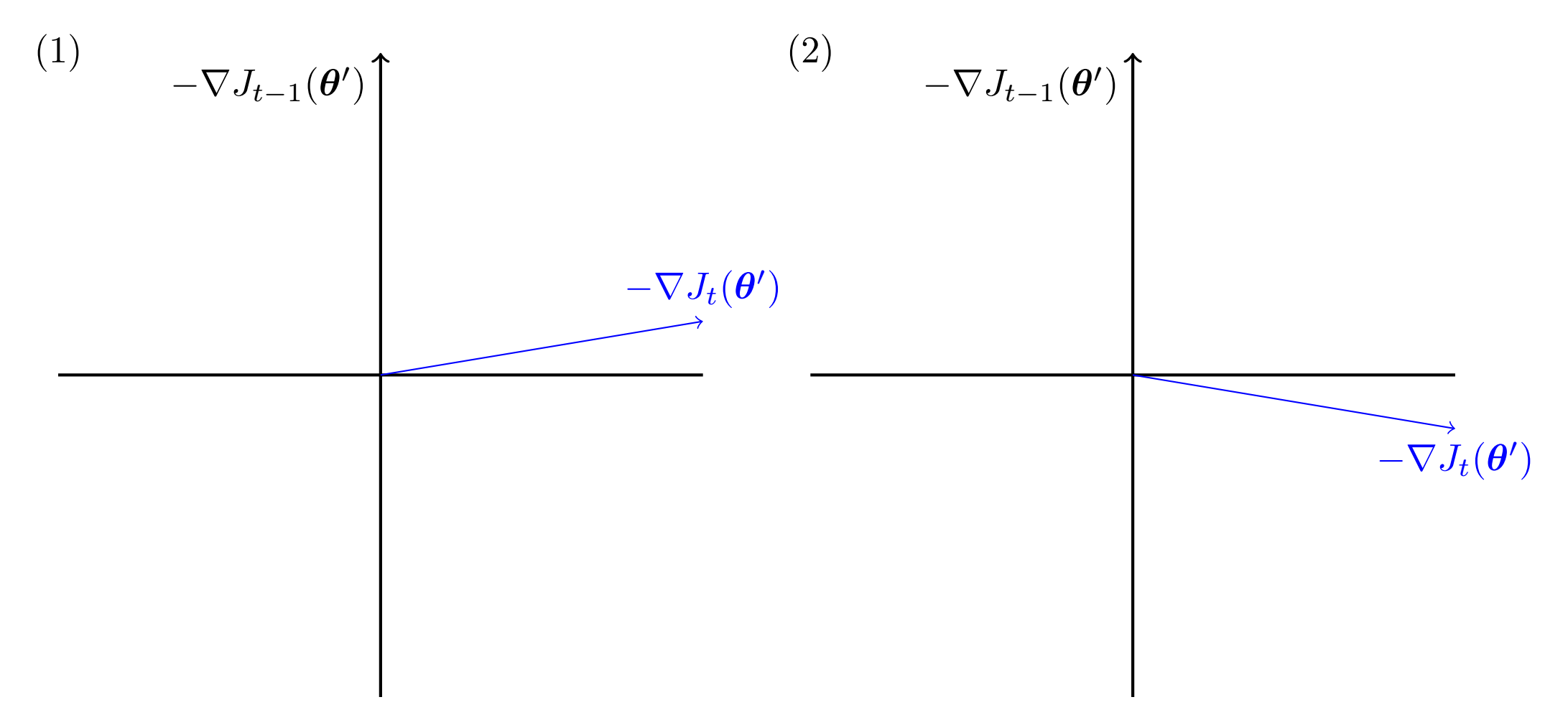}
\centering
\end{figure}

To address this, we propose a \emph{restricted optimum} (denoted by $\bm{\theta}_t^*$ at time $t$) as the tracking target of our algorithm.
At time $t$, the online player selects $\bm{\theta}_t$ based on observed $\{\nabla_1, ..., \nabla_{t-1}\}$.
As the network is trained using gradient descent, we propose to restrict the admissible weight set to the path formed from $\bm{\theta}_{t-1}^*$ and extending along the gradient vector $-\nabla_{t-1}$ (in other words, the path traversed by gradient descent).
The point $\bm{\theta}'$ along this path with the minimum $\norm{\nabla J_t(\bm{\theta}')}$ is the restricted optimum.
We argue that the trade-off between restricting the admissible weight space and solving the simplified problem is justified as other points in the weight space are not attainable via gradient descent and is thus unnecessary to consider all possible weight sets in $\Theta$.
Without assuming any time-varying dynamics, updating weights using an average of past gradients (similar to \citealp{Hazan:2017}) will induce a tracking error to the time-varying function.
To illustrate the restricted optimum concept, let $\bm{\theta}' = \bm{\theta}_{t-1}^*$ be our starting point of optimization, $\bm{g} = -\nabla J_{t-1}(\bm{\theta}')$ and $\bm{g}' = -\nabla J_{t}(\bm{\theta}')$. The possible scenarios during training are (also illustrated in Figure~\ref{fig:illustration_oes}):
\begin{enumerate}
    \item If $\left|cos^{-1} \frac{\left[\langle \bm{g}, \bm{g}'\rangle \right]}{\norm{\bm{g}}\norm{\bm{g}'}}\right| < \pi/2$, then moving along $\bm{g}$ will also improve $J_t(\bm{\theta}')$ until $\bm{g}$ is perpendicular to $\bm{g}'$ or $\bm{\theta}'$ has reached a local minima of $J_{t-1}$.
    \item If $\left|cos^{-1} \frac{\left[\langle \bm{g}, \bm{g}'\rangle \right]}{\norm{\bm{g}}\norm{\bm{g}'}}\right| \ge \pi/2$, then following $\bm{g}$ will not improve $J_t(\bm{\theta}')$ and training should terminate.
\end{enumerate}

\begin{figure}[H]
\caption{Illustration of estimating $\E\left[\norm{\bm{\theta}_t^* - \bm{\theta}_{t-1}^*}\right]$. Suppose $\bm{\theta}_t^* = [\theta_{1,t}^* \quad \theta_{2,t}^*]$ is a row vector with two elements. Twenty one random $\bm{\theta}_t^*$ vectors were drawn with each $\bm{\theta}_t^* - \bm{\theta}_{t-1}^*$ pair represented as an arrow. The circle has radius $\frac{1}{20} \sum_{t=2}^{21} \norm{\bm{\theta}_t^* - \bm{\theta}_{t-1}^*}$. $\bm{\theta}_t$ is regularized by limiting how far it can travel from $\bm{\theta}_{t-1}^*$ which is $\E\left[\norm{\bm{\theta}_t^* - \bm{\theta}_{t-1}^*}\right]$.}
\label{fig:illustration}
\includegraphics[width=12cm]{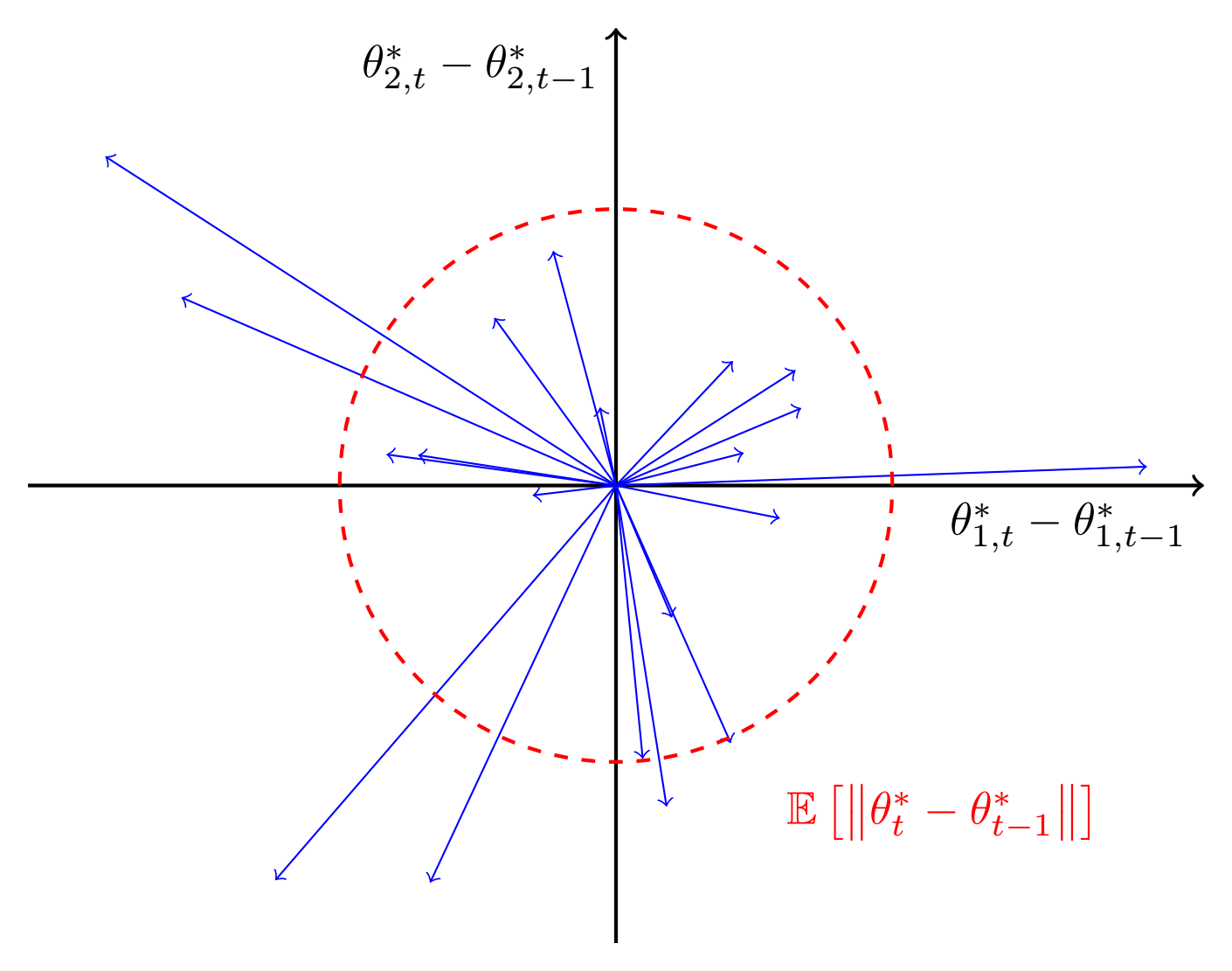}
\centering
\end{figure}

This observation motivates our online early stopping algorithm.
In this section, we will use $\bm{\theta}_t^*$ to denote restricted optimal weights at $t$ and $\bm{\theta}_t$ to denote the online player's choice of weights.
Suppose $\bm{\theta}_t^*$ evolves under the dynamics of,
\[ \bm{\theta}_t^* = \bm{\theta}_{t-1}^* - v_{t-1}\nabla J_{t-1}(\bm{\theta}_{t-1}^*), \]
where $v_{t-1}$ is sampled from an unknown distribution. $v_{t-1}$ can be interpreted as a \emph{regularizer} which provides the optimal prediction weights on $J_t$ if we are restricted to travelling along the direction of $-\nabla J_{t-1}(\bm{\theta}_{t-1}^*)$.
In this context, $\norm{\nabla J_t(\bm{\theta}_t^*)}$ is the minimum gradient suffered by the player.
Next, let $\tau_t^*$ be the optimal number of optimization steps at time $t$ and $\tau_{t}$ be the estimated number of optimization steps.
At iteration $t$, we solve optimal optimization steps $\tau_{t-2}^*$,
\begin{equation}
	\tau_{t-2}^* = \argmin_{\tau' \ge 0} J_{t-1}\left[\bm{\theta}_{t-2}^* - \eta \sum_{k=1}^{\tau'} \nabla J_{t-2}(\bm{\theta}_{t-2,k}^*)\right]. \label{eq:early_stop}
\end{equation}
We start from $t-2$ as solving $\tau_{t-1}^*$ requires $J_t$ which we are yet to observe.
This leads to optimal weights (the restricted optimum) trained on $J_{t-2}$ for prediction on $J_{t-1}$,
\begin{equation}
	\bm{\theta}_{t-1}^* = \bm{\theta}_{t-2}^* - \eta \sum_{k=1}^{\tau_{t-2}^*} \nabla J_{t-2}(\bm{\theta}_{t-2,k}^*), \label{eq:train_t-1}
\end{equation}
and can be approximated by,
\[ \bm{\theta}_{t-2}^* - \eta \sum_{k=1}^{\tau_{t-2}^*} \nabla J_{t-2}(\bm{\theta}_{t-2,k}^*) \approx \bm{\theta}_{t-2}^* - \eta \tau_{t-2}^* \nabla J_{t-2}(\bm{\theta}_{t-2}^*), \]
which implies $v_{t-2} \approx \eta \tau_{t-2}^*$.
To predict $\hat{\bm{r}}_t$, we choose $\tau_{t-1} = \frac{1}{t-2}\sum_{q=2}^{t-1} \tau_{t-q}^*$ and train \emph{prediction weights} on $J_{t-1}$ by substituting in $\floor{\tau_{t-1}+0.5}$ (the rounded up estimate of optimization steps),
\begin{equation}
	\bm{\theta}_t = \bm{\theta}_{t-1}^* - \eta \sum_{k=1}^{\floor{\tau_{t-1}+0.5}} \nabla J_{t-1}(\bm{\theta}_{t-1,k}^*) \approx \bm{\theta}_{t-1}^* - \eta \tau_{t-1} \nabla J_{t-1}(\bm{\theta}_{t-1}^*). \label{eq:theta_approx}
\end{equation}
As $\eta$ is a constant chosen by hyperparameter search, $\tau_{t-1}$ can be interpreted as a proxy to the regularizer $v_{t-1}$.
Using our $\beta$-smooth assumption (in Section~\ref{sec:definitions}) and substituting in definitions of $\bm{\theta}_t$ and $\bm{\theta}_t^*$ (in Equation~\ref{eq:theta_approx}), we obtain,
\begin{align}
    \norm{\nabla J_t(\bm{\theta}_t) - \nabla J_t(\bm{\theta}_t^*)} & \le \beta\norm{\bm{\theta}_t - \bm{\theta}_t^*}, \nonumber \\
    \sum_{t=2}^T \norm{\nabla J_t(\bm{\theta}_t) - \nabla J_t(\bm{\theta}_t^*)} & \le \sum_{t=2}^T \beta\norm{\bm{\theta}_t - \bm{\theta}_t^*}, \nonumber \\
    & \le \sum_{t=2}^T \beta\norm{\eta\tau_{t-1}^*\nabla J_{t-1}(\bm{\theta}_{t-1}^*) - \eta\tau_{t-1}\nabla J_{t-1}(\bm{\theta}_{t-1}^*)}, \label{eq:sum_gradients}
\end{align}
where we start from $t=2$ as our algorithm requires at least 2 cross-sectional observations.
The elegance of Equation~\ref{eq:sum_gradients} is that it conforms with the conventional notion of regret, with cumulative gradient deficit against an optimal outcome in place of cumulative loss.
As $\tau_{t-1}$ is the unbiased estimator of $\tau_{t-1}^*$, Equation~\ref{eq:sum_gradients} indicates that the cumulative deficit is asymptotically bounded by the variance of $\tau_{t-1}^*$.
This concept is illustrated in Figure~\ref{fig:illustration}.
If $\tau_{t-1}^*$ is constant, then $\tau_{t-1}$ will converge to $\tau_{t-1}^*$ and the optimal weights are achieved.
Conversely, if $\tau_{t-1}^*$ has high variance, then the player will suffer a larger cumulative gradient deficit.

\subsection{Online early stopping algorithm}

Our strategy is to modify the early stopping algorithm to recursively estimate $\tau_t$.
An outline is provided below as an introduction to the pseudocode in Algorithm~\ref{alg:OES}:
\begin{enumerate}
    \item At $t$, solve $\tau_{t-2}^*$ (Equation~\ref{eq:early_stop}) and $\bm{\theta}_{t-1}^*$ (Equation~\ref{eq:train_t-1}) by training on $J_{t-2}$ and validating against $J_{t-1}$ (line~\ref{lst:line:EarlyStop} of Algorithm~\ref{alg:OES}).
    \item Recursively estimate $\tau_{t-1}$ as the mean of observed $\{\tau_1^*, ..., \tau_{t-2}^*\}$ (line~\ref{lst:line:recursive_tau}).
    \item Start from $\bm{\theta}_{t-1}^*$ and perform gradient descent for $\floor{\tau_{t-1} + 0.5}$ iterations (Equation~\ref{eq:theta_approx}). The new weights are $\bm{\theta}_t$ (line~\ref{lst:line:start}--\ref{lst:line:end}).
    \item Predict using $\bm{\theta}_t$ (line~\ref{lst:line:predict}).
\end{enumerate}
$EarlyStopping$ on line~\ref{lst:line:EarlyStop} is outlined in Algorithm~\ref{alg:EarlyStopping}.
In our implementation of the algorithm, we have used stochastic gradient $\hat{\nabla}_{t-1}$ instead of the full gradient $\nabla_{t-1}$.
Algorithm~\ref{alg:EarlyStopping} contains the schematics of an early stopping algorithm with one modification adapted from
Algorithm~7.1 and Algorithm~7.2 in \cite{Goodfellow:2016}.
Validation is performed before the first training step to allow for the case where $\tau_{best} = 0$ (i.e., we start from the optimal weights).

\begin{algorithm}[H]
	\caption{General framework for online early stopping. The outer loop recursively estimates $\tau_{t-1}$.} \label{alg:OES}
	\begin{algorithmic}[1]
		\Require data $\bm{X}_t, \bm{r}_t \sim p_t$ at interval $t$; $\bm{\theta}_0^*$ initialized randomly
		\State $\tau' \gets 0$
		\For{$t = 2, ..., T$}
			\State $\tau', \bm{\theta}_{t-1}^* \gets$ \Call{EarlyStopping}{$\bm{\theta}_{t-2}^*, \bm{X}_{t-2}, \bm{r}_{t-2}, \bm{X}_{t-1}, \bm{r}_{t-1}$} \label{lst:line:EarlyStop}
			\State $\tau \gets \frac{\tau (t - 2) + \tau'}{t - 1}$ \label{lst:line:recursive_tau}
			\State $\bm{\theta} \gets \bm{\theta}_{t-1}^*$ \label{lst:line:start}
			\For{$i = 1, ..., \floor{\tau+0.5}$}
				\State $\bm{\theta} \gets \bm{\theta} - \eta \hat{\nabla}_{t-1}(\bm{\theta})$ \label{lst:line:train}
			\EndFor
			\State $\bm{\theta}_t \gets \bm{\theta}$ \label{lst:line:end}
			\State Receive input $\bm{X}_t$
			\State Predict $\hat{\bm{r}}_t \gets F(\bm{X}_t; \bm{\theta}_t)$ \label{lst:line:predict}
			\State Receive output $\bm{r}_t$
		\EndFor
	\end{algorithmic}
\end{algorithm}
\begin{algorithm}[H]
	\caption{Early stopping procedure. Training stops when validation loss does not improve by at least $\varepsilon$ for $Q$ iterations.} \label{alg:EarlyStopping}
	\begin{algorithmic}[1]
		\Require Maximum iterations $\{\Tau \in \N | \Tau > 0\}$; tolerance $\{\varepsilon \in \R | \varepsilon > 0\}$; patience $\{Q \in \N | Q > 0\}$; step size $\{\eta \in \R | \eta > 0\}$
		\Function{EarlyStopping}{$\bm{\theta}, \bm{X}_{train}, \bm{r}_{train}, \bm{X}_{test}, \bm{r}_{test}$}
			\State $\bm{\theta}_{best} \gets \bm{\theta}$
			\State $q \gets 0$
			\State $J_{best} \gets J(\bm{r}_{test}, F(\bm{X}_{test}; \bm{\theta}))$
			\For{$k = 1, ..., \Tau$}
				\State $\bm{\theta} \gets \bm{\theta} - \eta \hat{\nabla}J(\bm{r}_{train}, F(\bm{X}_{train}; \bm{\theta}))$
				\State $J' \gets J(\bm{r}_{test}, F(\bm{X}_{test}; \bm{\theta}))$
				\If{$J' < J_{best}$}
					\State $\tau_{best} \gets k$
					\State $\bm{\theta}_{best} \gets \bm{\theta}$
					\State $J_{best} \gets J'$
				\EndIf
				\If{$J'$ did not improve by at least $\varepsilon$}
					\State $q \gets q + 1$
					\If{$q \ge Q$}
						\State \textbf{break} \Comment{Assume convergence}
					\EndIf
				\Else
					\State $q \gets 0$
				\EndIf
			\EndFor
			\State \Return $\tau_{best}$, $\bm{\theta}_{best}$
		\EndFunction
	\end{algorithmic}
\end{algorithm}

In the next two sections, we conduct two empirical studies. First is based on simulation data which highlights the use of online early stopping, and the second on predicting U.S. stock returns based on the data set in \cite{Gu:2020} and is presented in Section~\ref{sec:us_equities}.

\section{Simulation study} \label{sec:simulation}

\subsection{Simulation data}

For the simulation study, we create the following synthetic data set:
\begin{itemize}
    \item $T = 180$ months, each month consists of $n = 200$ stocks.
    \item Each stock has $m = 100$ features, forming input matrix of $\bm{X} \in \R^{180 \times 200 \times 100}$ and output vector $\bm{r} \in \R^{180 \times 200}$.
    \item Let $x_{t,i,j}$ be the value of feature $j$ of stock $i$ at time $t$. Each feature value is randomly set to $x_{t,i,j} \sim N(0, 1)$.
    \item Each feature is associated with a latent factor $\psi_{t,j} = 0.95 \psi_{t-1,j} + 0.05 \delta_{t,j}$, where $\delta_{t,j} \sim N(0, 1)$ and $\psi_{0,j} \sim N(0, 1)$. $\psi_{t,j}$ follows a Wiener process and drifts over time.
    \item Each output value is $r_{t,i} = \sum_{j=1}^m \tanh(x_{t,i,j} \times \psi_{t,j}) + \epsilon_{t,i}$, where $\epsilon_{t,i} \sim N(0, 1)$. Thus, $\bm{r}_t$ is non-linear with respect to $\bm{X}_t$ and the relationship changes over time.
\end{itemize}
We have used the same network setup and hyperparameter ranges as the empirical study on U.S. equities (outlined in Table~\ref{tab:models}) but with a batch size of 50.
DNN has the same setup but is re-fitted at every 10-th time intervals.
The data set is split into three 60 interval blocks.
Hyperparameters for OES are chosen using a grid search, a procedure called \emph{hyperparameter tuning}.
For each hyperparameter combination, the network is trained on the first 60 intervals and validated on the next 60 intervals.
Hyperparameters with the minimum MSE in the validation set is used in the remaining 60 intervals as out-of-sample data.
Performance metrics are calculated using the out-of-sample set.
DTS-SGD follows the same training scheme as OES, with additional hyperparameters: window period $w \in \{5, 10, 20\}$ and forget factor $\alpha \in \{0.9, 0.8, 0.7\}$.

\subsection{Simulation results}

Our synthetic data requires the network to adapt to time-varying dynamics.
Table~\ref{tab:simulation_results} records results of the simulation.
DNN struggles to learn the time-varying relationships, with mean $R^2$ of \SI{-8.26}{\percent}
and mean rank correlation of \SI{-4.07}{\percent}.
This is expected as the expanding window approach used in DNN assumes the relationships at $t$ is best approximated by the average relationships in the observed past.
OES significantly outperforms the other two methods in this simple simulation, achieving mean $R^2$ of \SI{49.64}{\percent} and mean rank correlation of \SI{69.63}{\percent}.
This demonstrates OES's ability to track a non-linear, time-varying function reasonably closely.
There is a preference for higher $L_1$ regularization and learning rate.
In \cite{DTSSGD:2019}, the authors reported issues of exploding gradient with the \emph{static time-smoothed stochastic gradient descent} in \cite{Hazan:2017} and that DTS-SGD provided greater stability. 
In our simulation test, we observed gradient instability with DTS-SGD as well. During training, loss can increase after a weight update.
We hypothesize that a past gradient is taking network weights away from the direction of the current local minima and could be an issue with this general class of optimizers.
Lastly, we find that mean $R^2$ tends to be slightly lower than $R_{oos}^2$ (which is reasonable with a smaller denominator of a negative term).

\begin{table}[htbp]
\centering
\caption{Simulation results and selected hyperparameters by hyperparameter search averaged over time and ensemble networks. Values are in percentages unless specified ($w$, number of periods).}
\label{tab:simulation_results}
\begin{tabular}{@{} l c c c @{}}
\toprule
\% & DNN & OES & DTS-SGD \\
\midrule
Metrics & & & \\
\hspace{3mm} Pooled $R_{oos}^2$ & -7.12 & 50.22 & 0.13 \\
\hspace{3mm} Mean $R^2$ & -7.77 & 49.64 & -0.33 \\
\hspace{3mm} IC & -4.21 & 71.24 & 6.29 \\
Hyperparameters & & & \\
\hspace{3mm} Mean $L_1$ penalty & 0.01 & 0.09 & 0.04 \\
\hspace{3mm} Mean $\eta$ & 0.55 & 1.00 & 0.10 \\
\hspace{3mm} Mean $w$ (periods) & & & 14 \\
\hspace{3mm} Mean $\alpha$ & & & 83.00 \\
\bottomrule

\end{tabular}
\end{table}

\section{Predicting U.S. stock returns} \label{sec:us_equities}

\subsection{Model and U.S. equities data}

The U.S. equities data set in \cite{Gu:2020} consists of all stocks listed in NYSE, AMEX, and NASDAQ from March 1957 to December 2016.
Average number of stocks exceeds 5,200.
Excess returns over risk-free rate are calculated as forward one month stock returns over Treasury-bill rates.
As noted in Section~\ref{sec:definitions}, stock returns exhibit non-Gaussian characteristics.
Table~\ref{tab:return_describe} presents descriptive statistics of excess returns.
Monthly excess returns are positively skewed and contains possible outliers which may influence the regression.
We follow \cite{Gu:2020} in using MSE but note that MSE is not robust against outliers.
As noted in Section~\ref{sec:introduction}, we also provide an alternative setup which excludes microcap stocks.
The alternative setup and empirical results are presented in Section~\ref{sec:practical_sim}.

\begin{table}[htbp]
\small
\centering
\caption{Descriptive statistics of monthly excess returns of U.S. equities from April 1957 to December 2016, grouped into 10-Year periods.
Monthly excess returns appear to contain some extreme values, particularly on the positive end.
Variance of monthly excess returns varied over time.}
\label{tab:return_describe}
\begin{tabular}{@{} l c c c c c c @{}}
\toprule
\% & 1957-1966 & 1967-1976 & 1977-1986 & 1987-1996 & 1997-2006 & 2007-2016 \\
\midrule
Mean & 0.95 & 0.25 & 0.95 & 0.64 & 0.90 & 0.50 \\
Std Dev & 9.98 & 14.89 & 15.84 & 18.44 & 19.93 & 16.26 \\
Skew & 212.44 & 184.21 & 365.98 & 1059.88 & 502.41 & 783.70 \\
Min & -76.38 & -91.88 & -90.14 & -99.13 & -98.30 & -99.90 \\
1\% & -20.27 & -31.41 & -33.82 & -40.39 & -44.61 & -38.96 \\
10\% & -9.26 & -14.99 & -14.38 & -15.61 & -17.08 & -14.25 \\
25\% & -4.42 & -7.78 & -6.54 & -6.64 & -6.91 & -5.76 \\
50\% & -0.10 & -0.65 & -0.52 & -0.41 & 0.00 & 0.24 \\
75\% & 5.14 & 6.21 & 6.67 & 6.18 & 6.67 & 5.84 \\
90\% & 11.62 & 16.23 & 16.43 & 16.11 & 17.57 & 14.06 \\
99\% & 33.04 & 49.60 & 51.99 & 56.92 & 65.43 & 48.08 \\
Max & 255.29 & 432.89 & 1019.47 & 2399.66 & 1266.36 & 1598.45 \\
\bottomrule

\end{tabular}
\end{table}

Feature set includes 94 firm level features, 74 industry dummy
variables (based on first two digits of Standard Industrial Classification code, henceforth SIC) and interaction terms with 8 macroeconomic indicators.
The firm features and macroeconomic indicators used in \cite{Gu:2020} are based on \cite{Green:2017} and \cite{Welch:2008}, respectively.
Firm level features include share price based measures, valuation metrics and accounting ratios.
The purpose of interacting firm level features with macroeconomic indicators is to capture any time-varying dynamics that are related to
(common across all stocks) macroeconomic indicators. For instance, suppose valuation metrics have a stronger relationship with stock returns during periods of high
inflation. Then, this information will be encoded in the interaction term.
The aggregated data set therefore contains $94 \times (8 + 1) + 74 = 920$ features.
Each feature has been appropriately lagged to avoid look-forward bias, and is cross-sectionally ranked and scaled to $[-1, 1]$.
Table~A.6 in the Internet Appendix of \cite{Gu:2020} contains the full list of firm features.

A subset of the data is available on Dacheng Xiu's website\footnote{Dacheng Xiu's website \url{https://dachxiu.chicagobooth.edu/}}
which contains 94 firm level characteristics and 74 industry classification.
Our main result uses $94 + 74 = 168$ firm level features but results with the full 920 features are also provided as a comparison.
At this point, it is useful to remind readers that our goal
is to track a time-varying function when the time-varying dynamics are unknown. In other words, we assume that time-varying dynamics between
stock returns and features are not well understood or are unobservable.
As such, the subset of data without interaction terms is sufficient for our problem.
If macroeconomic indicators do encode time-varying dynamics,
our network will track changing macroeconomic conditions automatically.

Data is divided into 18 years of training (from 1957 to 1974), 12 years of validation (1975-1986), and 30 years of out-of-sample tests (1987-2016).
We use monthly total returns of individual stocks
from CRSP. Where stock price is unavailable at the end of month, we use the last available price during the month.
Table~\ref{tab:models} records test configurations as outlined in \cite{Gu:2020} and in our replication. A total of six hyperparameter combinations ($L_1$ penalty and $\eta$ in Table~\ref{tab:models}) are tested.
We use the same training scheme as \cite{Gu:2020} to train DNN. Once hyperparameters are tuned, the same network is used to make predictions in the out-of-sample set for 12 months.
Training and validation sets are rolled forward by 12 months at the end of every December and the model is re-fitted.
An ensemble of 10 networks is used, where each prediction $r_{t,i}$ is the average prediction of 10 networks.

\begin{table}[htbp]
\centering
\setlength\tabcolsep{2pt}
\caption{Disclosed model parameters in \cite{Gu:2020} and in our replication. We fill missing values with the cross-sectional median or zero if median is unavailable.
`H' is hidden layer activation. `O' is output layer activation.
ADAM is the optimizer proposed by \cite{ADAM:2015}.}
\label{tab:models}
\begin{tabular}{@{} l c c @{}}
\toprule
Parameter & \cite{Gu:2020} & This paper \\
\midrule
Preprocessing & Rank [-1, 1]; Fill median & Rank [-1, 1]; Fill median/0  \\
Hidden layers & 32-16-8 & 32-16-8 \\
Activation & H: ReLU / O: Linear & H: ReLU / O: Linear \\
Batch size & 10,000 & DNN 10,000 / OES 1,000 \\
Batch normalization & Yes & Yes \\
$L_1$ penalty & $[10^{-5}, 10^{-3}]$ & $\{10^{-5}, 10^{-4}, 10^{-3}\}$ \\
Early stopping & Patience 5 & Patience 5 / Tolerance 0.001 \\
Learning rate $\eta$ & $[0.001, 0.01]$ & $\{0.001, 0.01\}$ \\
Optimizer & ADAM & ADAM \\
Loss function & MSE & MSE \\
Ensemble & Average over 10 & Average over 10 \\
\bottomrule

\end{tabular}
\end{table}

To train OES, we keep the first 18 years (to 1974) as training data and next 12 years (to 1986) as validation data.
For each permutation of hyperparameter set, we have trained an online learner up to 1986.
Hyperparameter tuning is only performed once on this period, as opposed to every year in \cite{Gu:2020}.
As the algorithm
does not depend on a separate set of data for validation, we simply take the hyperparameter set with the lowest
monthly average MSE over 1975-1986 as the best configuration to use for rest of the data set.
Batch size of 1,000 for OES was chosen arbitrarily.

\subsection{Predicting U.S. stock returns} \label{sec:us_results}

In this section, we present our U.S. stock return prediction results.
DTS-SGD did not complete training with a reasonable range of hyperparameters due to exploding gradient and is omitted from this section.
As an overarching comment,
$R^2$ for both DNN and OES on U.S. stock returns are very low, and are consistent with the findings of \cite{Gu:2020}.
First, results with and without interaction terms are presented in Table~\ref{tab:initial_results}, keeping in mind that our method should be compared against DNN without interaction terms.
Without interaction terms, OES and DNN achieve IC of \SI{4.53}{\percent} and \SI{3.82}{\percent}, respectively.
The relatively high correlation of OES (compared to DNN) indicates that it is better at differentiating relative performance between stocks.
This is particularly important in our use case as practitioners build portfolios based on expected relative performance of stocks.
For instance, a long-short investor will buy top ranked stocks and short sell bottom ranked stocks and earn the difference in relative return between the two baskets of stocks.
Mean $R^2$ are \SI{-12.14}{\percent} and \SI{-9.68}{\percent} for OES and DNN, respectively.
Note that the denominator of mean $R^2$ is adjusted by the cross-sectional mean of excess returns.
Therefore, negative means $R^2$ of both OES and DNN indicate that neither method is able to accurately predict the magnitude of cross-sectional returns.
Finally, OES scores \SI{-2.48}{\percent} on $R_{oos}^2$ and DNN scores \SI{0.22}{\percent}.
The low values of both methods underscore the difficulty in return forecasting.
DNN achieves higher Sharpe ratio than OES, at 1.63 and 0.83, respectively.
As we will point out in Section~\ref{sec:practical_sim}, the high Sharpe ratio of DNN is driven by microcap stocks.
Despite the very low $R^2$, both methods are able to generate economically meaningful returns.
This underscores our argument that $R^2$ is not the best measure of performance and verifies practitioners' choice of correlation as the preferred measure.
We observed similar performance with interaction terms, suggesting that the 8 macroeconomic time series have little interaction effect with the 94 features.
In the subsequent results in this section, we only report statistics without interaction terms.

\begin{table}[htbp]
\centering
\caption{Predictive performance on U.S. equities. Pooled $R_{oos}^2$ is calculated across the entire out-of-sample period as a whole. Mean $R^2$ and IC are calculated cross-sectionally for each month then averaged across time.
P10-1 is the average monthly spread between top and bottom deciles.
Sharpe ratio is based on P10-1 return spread and annualized
Mean hyperparameters are calculated over the ensemble of 10 networks and across all periods. \emph{As reported} are results in \cite{Gu:2020}.}
\label{tab:initial_results}
\begin{tabular}{@{} l c c c c c c @{}}
\toprule
 & \multicolumn{3}{c}{With Interactions} && \multicolumn{2}{c}{W/O Interactions} \\
\cmidrule(l){2-4} \cmidrule(l){6-7}
\% & As reported & DNN & OES && DNN & OES \\
\midrule
Metrics & & & & & & \\
\hspace{3mm} Pooled $R_{oos}^2$ & 0.4 & 0.13 & -1.93 & & 0.22 & -2.48 \\
\hspace{3mm} Mean $R^2$ & & -9.89 & -11.93 & & -9.68 & -12.17 \\
\hspace{3mm} IC & & 3.51 & 4.22 & & 3.82 & 4.53 \\
\hspace{3mm} P10-1 & 3.27 & 1.83 & 2.10 & & 2.39 & 2.41 \\
\hspace{3mm} Sharpe ratio & 2.36 & 0.94 & 0.72 & & 1.63 & 0.83 \\
Hyperparameters & & & & & & \\
\hspace{3mm} Mean $L_1$ penalty & & 0.0012 & 0.0154 & & 0.0024 & 0.0028 \\
\hspace{3mm} Mean $\eta$ & & 0.77 & 0.10 & & 0.67 & 0.10 \\
\bottomrule

\end{tabular}
\end{table}

\begin{table}[htbp]
\centering \small
\caption{Predicted and realized mean returns by decile where each row represents a decile. \emph{P1} is the mean excess returns of the first decile (0-10\% of bottom ranked stocks) and \emph{P10-1} is \emph{P10} less \emph{P1} showing the return spread between the best decile relative to the worst decile. \emph{As reported} are original results from Table~A.9 in \cite{Gu:2020}.}
\label{tab:initial_portfolios}
\begin{tabular}{@{} l c c c c c c c c @{}}
\toprule
 & \multicolumn{2}{c}{As reported} && \multicolumn{2}{c}{DNN} && \multicolumn{2}{c}{OES} \\
\cmidrule{2-3} \cmidrule{5-6} \cmidrule{8-9}
\% & Predicted & Realized && Predicted & Realized && Predicted & Realized \\
\midrule
P1 & -0.31 & -0.92 && -0.59 & -0.47 && -3.53 & -0.50 \\
P2 & 0.22 & 0.16 && 0.09 & 0.15 && -1.96 & 0.03 \\
P3 & 0.45 & 0.44 && 0.37 & 0.54 && -1.07 & 0.27 \\
P4 & 0.60 & 0.66 && 0.55 & 0.64 && -0.34 & 0.48 \\
P5 & 0.73 & 0.77 && 0.70 & 0.73 && 0.30 & 0.67 \\
P6 & 0.85 & 0.81 && 0.84 & 0.78 && 0.88 & 0.85 \\
P7 & 0.97 & 0.86 && 0.99 & 0.85 && 1.46 & 1.04 \\
P8 & 1.12 & 0.93 && 1.17 & 0.96 && 2.10 & 1.18 \\
P9 & 1.38 & 1.18 && 1.43 & 1.26 && 2.89 & 1.42 \\
P10 & 2.28 & 2.35 && 2.33 & 1.92 && 4.25 & 1.91 \\
P10-1 & 2.58 & 3.27 && 2.92 & 2.39 && 7.78 & 2.41 \\
\bottomrule

\end{tabular}
\end{table}

\begin{figure}[htbp]
\caption{Cumulative mean excess returns by decile sorted based on predictions by DNN and OES. Each portfolio follows the same construction as described in Table~\ref{tab:initial_portfolios}. However, cumulative sum of mean excess returns of each portfolio is presented in the chart.}
\label{fig:fractiles}
\includegraphics[width=12cm]{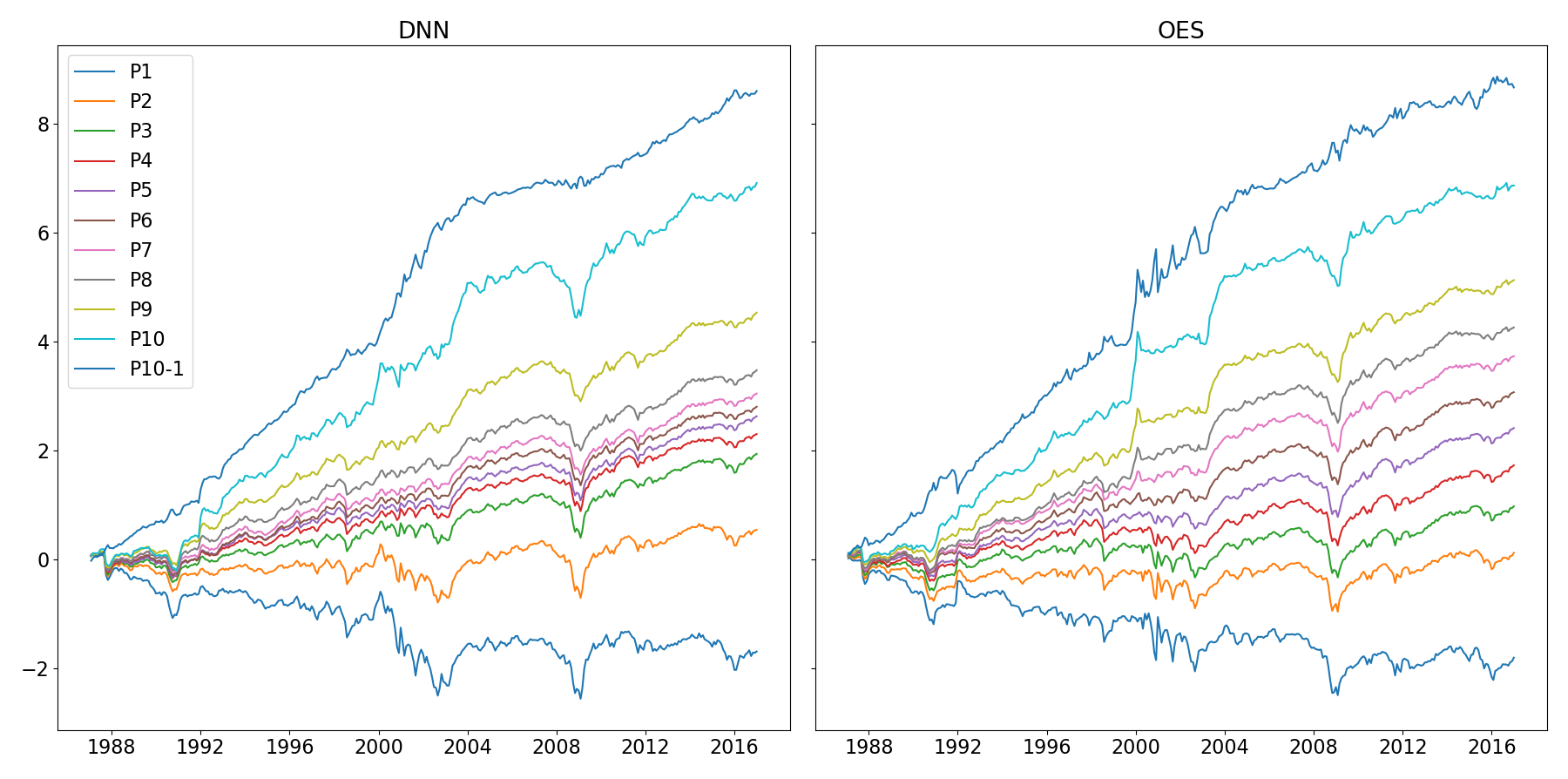}
\centering
\end{figure}

So why do IC and $R_{oos}^2$ diverge? The answer lies in Table~\ref{tab:initial_portfolios} and Figure~\ref{fig:fractiles}. In here, we form decile
portfolios based on predicted returns over the next month and track their respective realized returns.
OES predicted values span a wider range than DNN.
This has contributed to a lower $R^2$, even though OES is able to better differentiate relative performance between stocks.
DNN used a pooled data set which will average out time-varying effects. As a result, the average gradient will likely
smaller in magnitude. This is evident from the lower mean $L_1$ penalty and higher learning rate $\eta$ chosen by validation.
By contrast, OES trains on each time period individually and the norm of the gradient presented to
the network at each period is likely to be larger.
This led to a lower learning rate chosen by validation.
Hence, variance of OES predicted values is higher and potentially requires higher or different forms of regularization.

In Table~\ref{tab:initial_portfolios} and Figure~\ref{fig:fractiles}, we observe that the prediction performance of DNN
is concentrated on the extremities, namely P1 and P10, with realized mean returns of \SI{-0.47}{\percent} and
\SI{1.92}{\percent} respectively.
Stocks between P3 and P7 are not well differentiated. By contrast, OES is better at ranking stocks across the entire spectrum.
Realized mean returns of OES are more evenly spread across the deciles, resulting
in higher correlation than DNN.
P10-1 realized portfolio returns are similar across DNN and OES at \SI{2.39}{\percent} and \SI{2.41}{\percent}, respectively.
However, the difference in mean return spread increases when calculated on a quintile basis (mean return of top \SI{20}{\percent} of stocks minus bottom \SI{20}{\percent}), to \SI{1.75}{\percent} and \SI{1.90}{\percent} for DNN and OES, respectively.
This reflects better predictiveness in the middle of the spectrum of OES.
An investor holding a well diversified portfolio is more likely to utilize predictions closer to the center of the distribution and experience relative returns that are reminiscent of the quintile spreads (and even tertile spreads) rather decile spreads.

\subsection{Time-varying feature importance}

So far, our tests are predicated on time-varying relationships between features and stock returns.
How do features' importance change over time?
To examine this, at every time period we train the OES model and make a baseline prediction.
For each feature $j = 1, ..., m$, all values of $j$ are set to zero and a new prediction is made.
A new $R^2$ is calculated between the new prediction and the baseline prediction, denoted as $R_{t,J}^2$.
Importance of feature $j$ at time $t$ is calculated as ${FI}_{t,j} = 1 - R_{t,j}^2$.
Our measure tracks features that the network is using.
This is different to the procedure in \cite{Gu:2020} where $R^2$ is calculated against actual stock returns, rather than a baseline prediction. 

\begin{figure}[htbp]
\caption{Top 5 features based on rolling 12-month average feature importance over 1987-1991. Three rapid falls can be seen which coincide with the 1990-91 U.S. recession, Dot-com bubble (2000-03) and the Global Financial Crisis (2007-09). These periods are shaded for reference.}
\label{fig:abs_r2_delta}
\includegraphics[width=12cm]{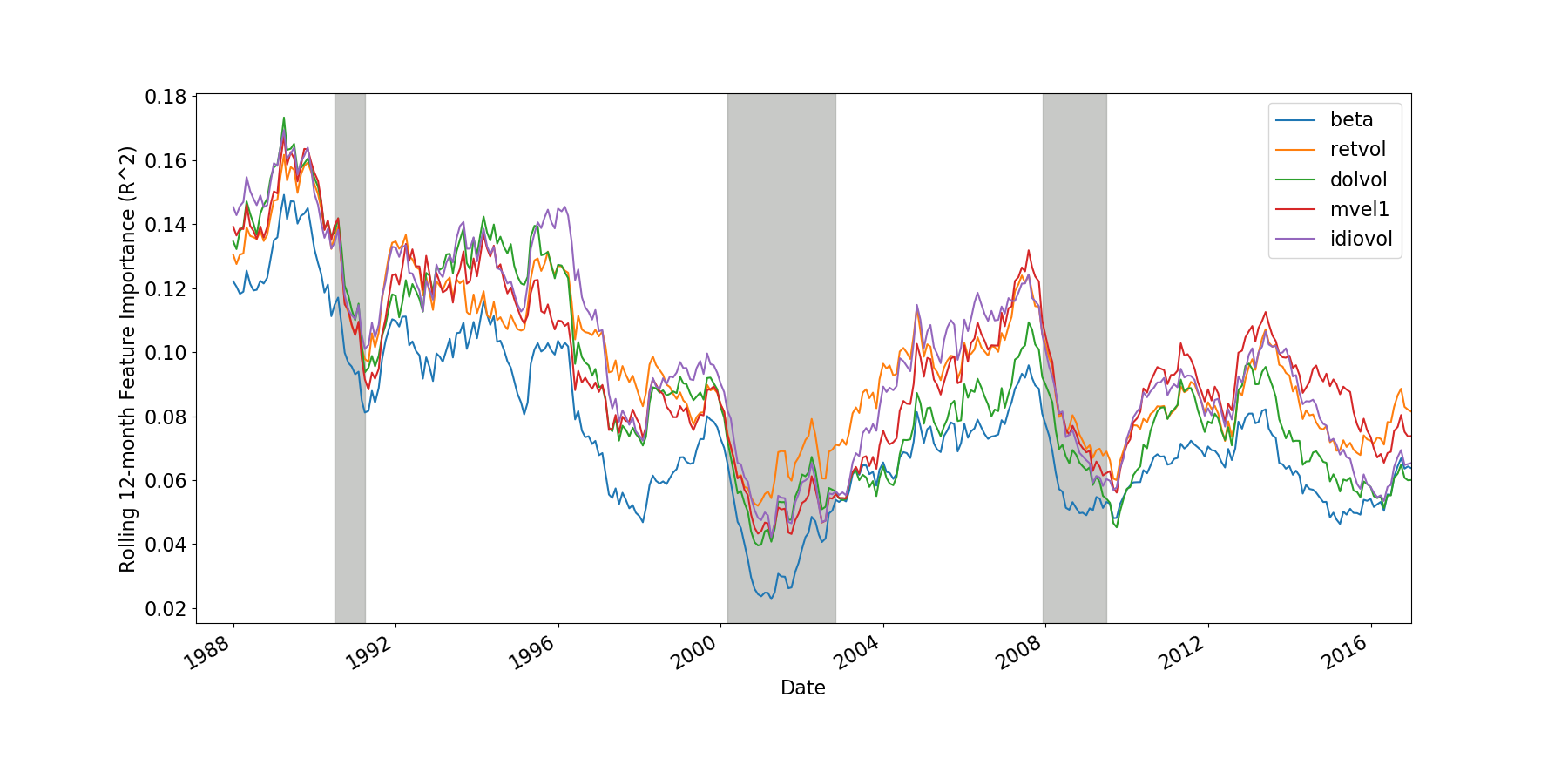}
\centering
\end{figure}

To illustrate the inadequacy of a non-time-varying model, we first track feature importance over January 1987 to December 1991.
The top 10 features with the highest feature importance are (in order of decreasing importance): \emph{idiovol} (CAPM residual volatility),
\emph{mvel1} (log market capitalization), \emph{dolvol} (monthly traded value), \emph{retvol} (return volatility),
\emph{beta} (CAPM beta), \emph{mom12m} (12-month minus 1-month price momentum), \emph{betasq} (CAPM beta squared),
\emph{mom6m} (6-month minus 1-month month price momentum), \emph{ill} (illiquidity), and \emph{maxret} (30-day max daily return).
Rolling 12-month averages were calculated to provide a more discernible trend, with the top 5 shown in Figure~\ref{fig:abs_r2_delta}.
Feature importance exhibits strong time-variability.
Rolling 12-month average feature importance fell from 14-16\% at the start of the out-of-sample period to a trough of 2-6\% before rebounding.
This indicates that the network would have changed considerably over time.
Rapid falls in feature importance can be seen in Figure~\ref{fig:abs_r2_delta}, over 1990-91, 2000-01 and 2008-09.
These periods correspond to the U.S. recession in early 1990s, the Dot-com bubble and the Global Financial Crisis, respectively.
Thus, market distress may explain rapid changes in feature importance.

\begin{figure}[H]
\caption{Yearly average $R^2$ to baseline predictions (in decimal). The OES network appeared to use only a handful of features. Shades of feature importance are distinctly lighter over 2000-02, 2008-09, and to a lesser extent in 1990 and 1997. Importance of some features have eroded over time (e.g. \emph{dolvol}, \emph{maxret} and \emph{turn}).}
\label{fig:abs_r2_trial43}
\includegraphics[width=15cm]{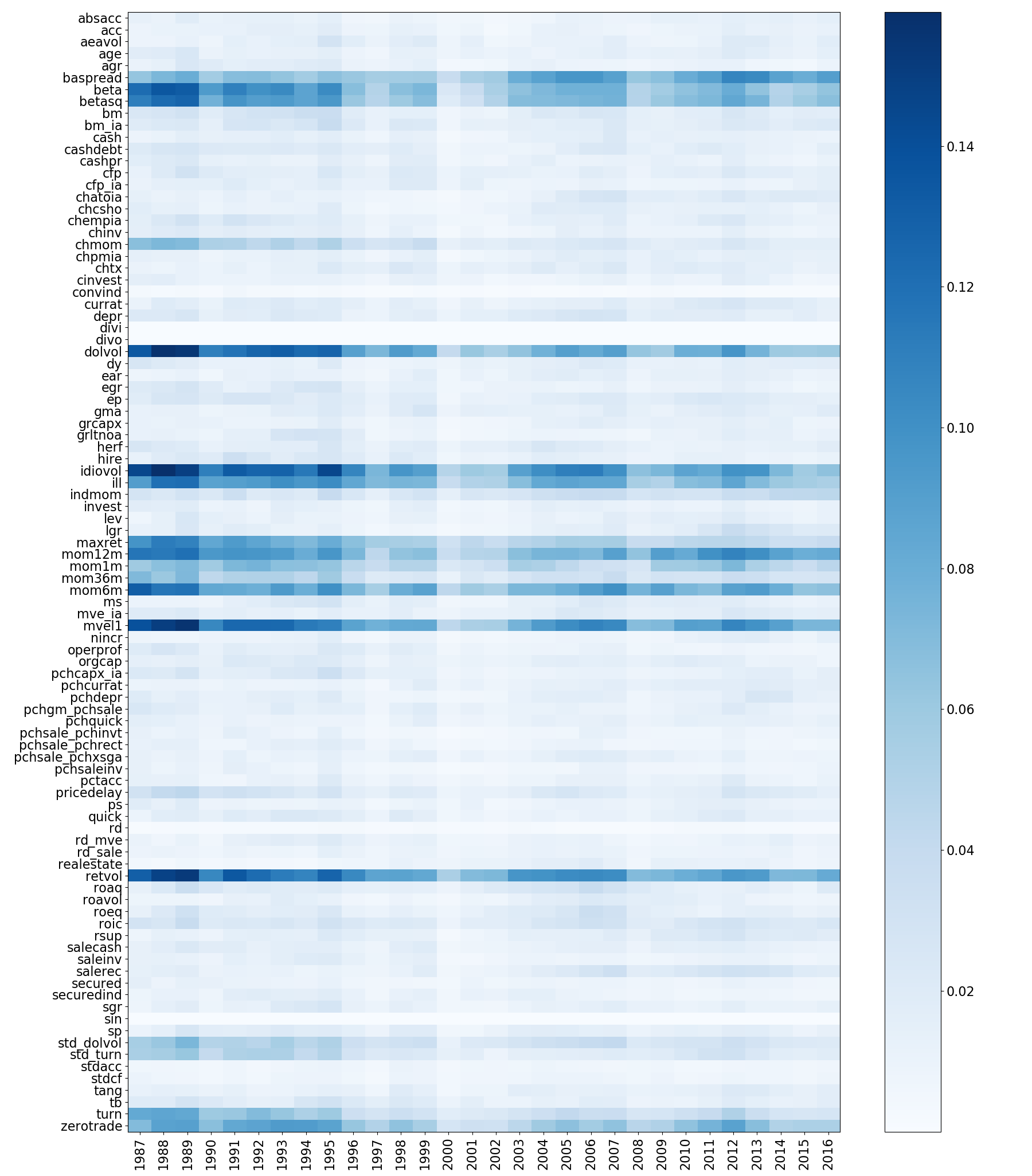}
\centering
\end{figure}

\begin{figure}[H]
\caption{Rolling 12-month average $R^2$ to baseline prediction of SIC code 13, 60 and 73, as proxies for oil \& gas companies, banks and technology companies, respectively. $R^2$ of technology companies peaks over 2001-02, banks over 2008-10, and oil companies over 2015-16. Duration of 1990-91 U.S. recession, Dot-com bubble, Global Financial Crisis and the 2014-16 oil glut have been shaded in grey.}
\label{fig:bank_it_oil_r2}
\includegraphics[width=12cm]{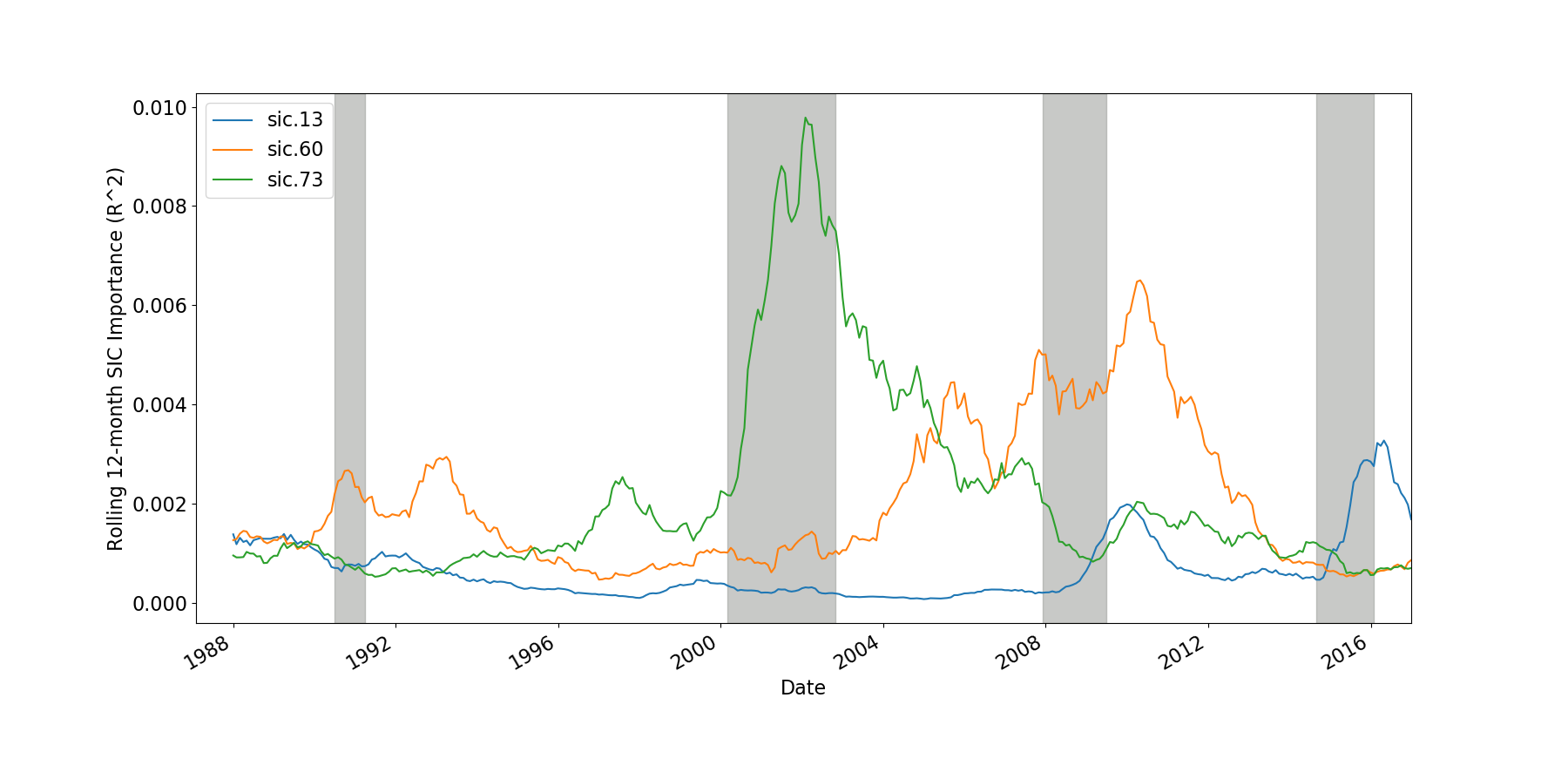}
\centering
\end{figure}

Next, we examine changes in importance for all features on a yearly basis.
Figure~\ref{fig:abs_r2_trial43} displays considerable year-to-year variations in feature importance.
As there are just a few clusters of features with relatively higher feature importance, the network's predictions can be attributed to a small set of features.
This is likely due to the use of $L_1$ regularization which encourages sparsity.
There is an overall trend towards lower importance over time, consistent with publication-informed trading hypothesis of \cite{AcademicDestroy:2016}.
For instance, the importance of the market capitalization (\emph{mvel1}) has decreased over time, as documented in \citep{Horowitz:2000}.
There are periods of visibly lower importance for all features, over 2000-02 and 2008-09, and to a lesser extent 1990 and 1997 (Asian financial crisis).
If all features have lower importance during market distress, then what explains stock returns during these periods?
To answer this question, we turn to importance of sectors, using SIC 13 (Oil and Gas), 60 (Depository Institutions) and 73 (Business Services) as proxies for oil companies, banks and technology companies, respectively.
Figure~\ref{fig:bank_it_oil_r2} records the rolling 12-month average $R^2$ to baseline prediction of banks, oil and technology companies.
The peak of importance of SIC 73 overlaps with the Dot-com bubble and peak of SIC 60 occurs just after the Global Financial Crisis (which started as a sub-prime mortgage crisis).
Importance of SIC 13 peaked in 2016, coinciding with the 2014-16 oil glut which saw oil prices fell from over US\$100 per barrel to below US\$30 per barrel. This is an example of how an exogenous event that is confined to a specific industry impacts on predictability of stock returns.
Thus, a plausible explanation for the observed results is that firm features explain less of cross-sectional returns during market shocks, which becomes increasingly explained by industry groups.
This is particularly true if the market shock is industry related.
For instance, technology companies during the Dot-com bubble, oil companies during an oil crisis and lodging companies during a pandemic.
This underscores the importance to have a dynamic model that adapts to changes in the true model.

\subsection{Investable simulation} \label{sec:practical_sim}

As noted in Section~\ref{sec:introduction}, the data set in \cite{Gu:2020} contains many stocks that are small and illiquid.
The \cite{SECMicro:2013} defines ``microcap'' stocks as companies with market capitalization below US\$250--300 million and ``nanocap'' stocks as companies with market capitalization below US\$50 million.
At the end of 2016, there are over to 1,300 stocks with market capitalization below US\$50 million and over 1,800 stocks with market capitalization between US\$50 million and US\$300 million.
Together, microcap and nanocap stocks constitute close to half of the data set as at 2016.
Thus, we also provide results excluding these difficult to trade stocks.
At the end of every June, we calculate breakpoint based on the 5-th percentile of NYSE listed stocks and exclude stocks with market capitalization below this value.
Once rebalanced, the same set of stocks are carried forward until the next rebalance (unless the stock ceases to exist).
This cutoff is chosen to approximately include the larger half of U.S.-listed stocks, with the average number of stocks exceeding 2,600.
We label this data set as the \emph{investable set}.
To mitigate the impact of outliers, we also winsorize excess returns at \SI{1}{\percent} and \SI{99}{\percent} for each month (separately).
Winsorized returns are then standardized by subtracting the cross-sectional mean and dividing by cross-sectional standard deviation.
Standardization is a common procedure in machine learning and can assists in network training \citep{LeCun:2012}.
Predicting a dependent variable with zero mean also removes the need to predict market returns which is embedded in stocks' excess returns (over risk free rate).
This transformation allows the neural network to more easily learn the relationships between relative returns and firm characteristics.

\begin{table}[htbp]
\centering
\caption{Predictive performance on the investable set. Ensemble is the average of standardized predictions of the two methods. Pooled $R_{oos}^2$ is calculated across the entire out-of-sample period as a whole. Mean $R^2$ and IC are calculated cross-sectionally for each month then averaged across time.
P10-1 is the average monthly spread between top and bottom deciles.
Sharpe ratio is based on P10-1 return spread and annualized.
Mean hyperparameters are calculated over the ensemble of 10 networks and across all periods.}
\label{tab:practical_results}
\begin{tabular}{@{} l c c c @{}}
\toprule
\% & DNN & OES & Ensemble \\
\midrule
Metrics & & & \\
\hspace{3mm} Pooled $R_{oos}^2$ & 0.35 & -1.37 & \\
\hspace{3mm} Mean $R^2$ & 0.35 & -1.37 & \\
\hspace{3mm} IC & 5.74 & 6.05 & 6.29 \\
\hspace{3mm} P10-1 & 1.69 & 2.41 & 2.60 \\
\hspace{3mm} Sharpe ratio & 0.69 & 0.82 & 0.96 \\
Hyperparameters & & & \\
\hspace{3mm} Mean $L_1$ penalty & 0.0211 & 0.0046 & \\
\hspace{3mm} Mean $\eta$ & 0.87 & 0.10 & \\
\bottomrule

\end{tabular}
\end{table}

\begin{figure}[htbp]
\caption{Monthly and rolling 12-month correlation between predictions of OES and DNN. Duration of 1990-91 U.S. recession, Dot-com bubble, Global Financial Crisis and the 2014-16 oil glut have been shaded in grey.}
\label{fig:oes_dnn_corr}
\includegraphics[width=12cm]{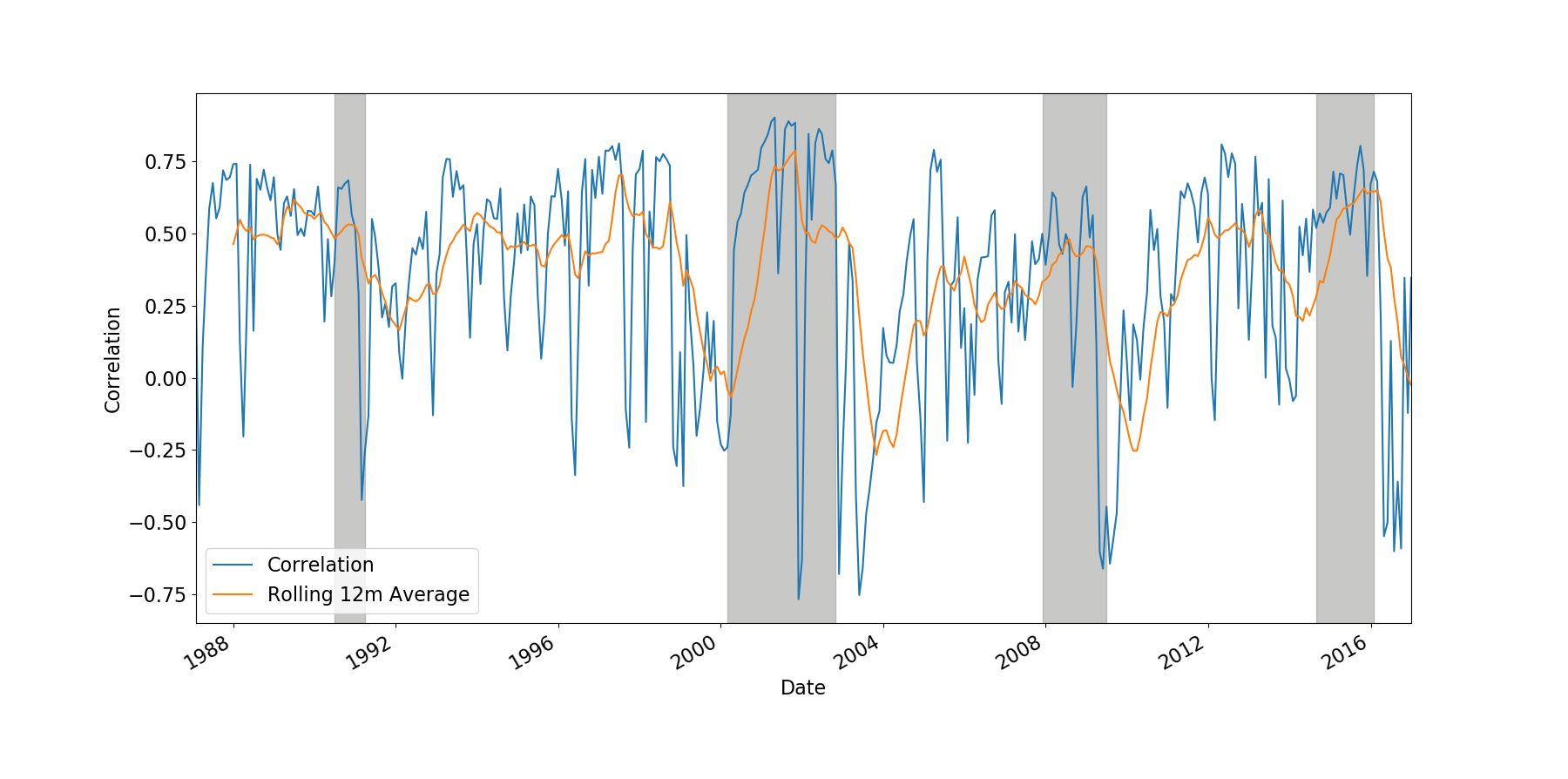}
\centering
\end{figure}

Results based on this investable set are presented in Table~\ref{tab:practical_results}.
Both $R_{oos}^2$ and IC improved once microcaps are excluded, with OES scoring \SI{6.05}{\percent} on IC and DNN on \SI{5.74}{\percent}.
However, DNN experienced a significant drop in mean decile spread (to \SI{1.69}{\percent} per month) and Sharpe ratio (0.69), suggesting that microcaps are significant contributors to the results using the full data set.
By contrast, mean decile spread and Sharpe ratio remain stable for OES, at \SI{2.41}{\percent} and 0.82, respectively.
This indicates that predictive performance of OES was not driven by microcap stocks.
We believe this is a meaningful result for practitioners as this subset represents a relatively accessible segment of the market for institutional investors.
An ensemble based on the average of cross-sectionally standardized predictions of the two models achieved the best IC, decile spread and Sharpe ratio relative to OES and DNN.
Mean monthly correlation between OES and DNN is only \SI{35.9}{\percent}.
Thus, an ensemble based on the two methods can effectively reduce variance of the predictions.
Monthly correlations between the two models are presented in Figure~\ref{fig:oes_dnn_corr}.
We observe that correlation tends to be lowest immediately after a recession or crisis.
We hypothesise that OES is quicker to react to an economic recovery.

\begin{figure}[htbp]
\caption{Cumulative mean excess returns by decile sorted based on predictions by DNN and OES in the investable set.}
\label{fig:fractiles_investable}
\includegraphics[width=12cm]{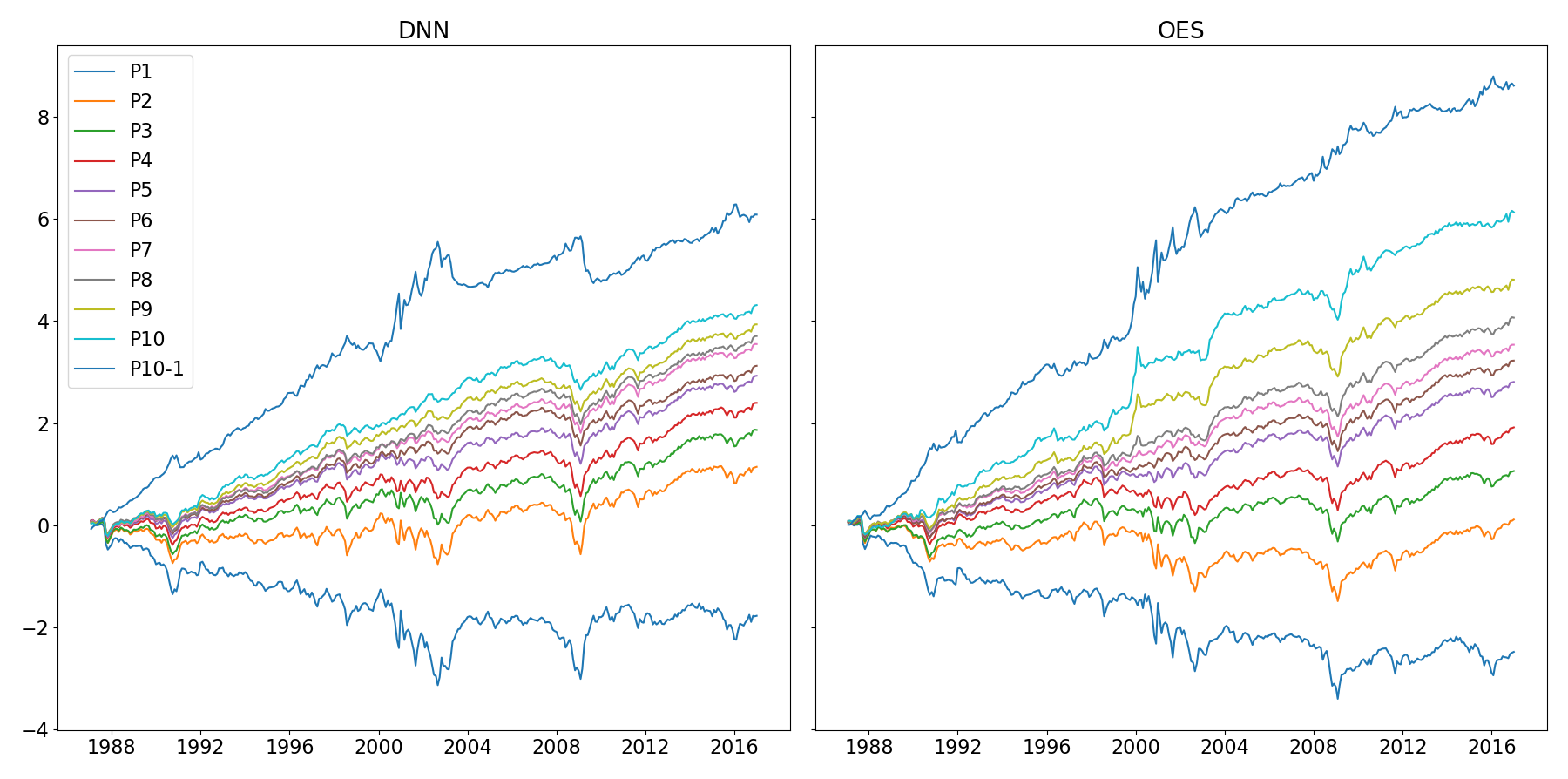}
\centering
\end{figure}

Turning to cumulative decile returns presented in Figure~\ref{fig:fractiles_investable}, we observe significant drawdowns for DNN during recovery phases of the Dot-com bubble and Global Financial Crisis.
P1 of DNN bounced back sharply during these episodes, causing sharp drops in decile spreads and are consistent with \emph{momentum crashes} \citep{MomCrash:2016}.
By contrast, decile spreads of OES appear to react to the recovery more quickly.
Consistent with prior findings, the spreads between decile 3 to 7 are also better under OES than DNN in the investable set.
Given these favorable characteristics, practitioners are likely to find OES a useful tool to add to the armory of prediction models.

\section{Conclusions} \label{sec:conclusion}

Stock return prediction is an arduous task. The true model is noisy, complex and time-varying.
Mainstream deep learning research has focused on problems that do not vary over time and, arguably, time-varying applications
have seen less advancements. In this work, we propose an online early stopping algorithm that is easy to implement
and can be applied to an existing network setup.
We show that a network trained with OES can
track a time-varying function and achieve superior performance to DTS-SGD, a recently proposed online non-convex optimization technique.
Our method is also significantly faster, as only two periods
of training data are required at each iteration, compared to the pooled method used in \cite{Gu:2020} which re-trains
the network on the entire data set annually. In our tests, the pooled method took \SI{5.5}{hours} to iterate through
the entire data set (an ensemble of ten networks therefore takes \SI{55}{hours})\footnote{Tests performed on
AMD Ryzen\texttrademark~7 3700X, Python 3.7.3, Tensorflow 1.12.0 and Keras 2.2.4. Hyperparameter grid search
was performed concurrently.}.
By contrast, our method took \SI{44.25}{mins} for a single pass over the entire data set (an ensemble of ten networks took \SI{7.4}{hours}).

\cite{Gu:2020} suggested that a small data set and low signal-to-noise ratio were reasons for the lack of improvement with
a deeper network. To this end, we show that only a handful of features contribute to predictive performance.
This may be due to correlation between features and the use of $L_1$ regularization which encourages sparsity.
We also find evidence of time-varying feature importance.
In particular, features such as log market capitalization (the size effect) and 12-month minus 1-month momentum have seen a gradual decrease to their importance towards the end of our test period, consistent with the publication-informed trading hypothesis of \cite{AcademicDestroy:2016}.
We find that sectors can also exhibit time-varying importance (for instance, technology stocks during the Dot-com bubble).
These results have strong implications for practitioners forecasting stock returns using well known asset pricing anomalies.
Excluding microcaps, we find that OES offers superior predictive performance in a subuniverse that is accessible to institutional investors.
We find that correlation between OES and DNN is at its lowest after a recession or crisis.
We argue that this is driven by faster reactions of OES in tracking the recovery.
An ensemble based on the average prediction of the two models achieves the best IC and Sharpe ratio, suggesting that the two methods may be complementary.

From an academic perspective, recent advances in deep learning such as dropout and \emph{residual connections} \citep{ResNet:2016} may allow deeper networks to be trained, which enables more expressive asset pricing models.
Given the higher variance of predictions produced by OES, future work should explore alternative methods of regularization including dropouts, $L_2$ penalty or a mixture of regularization techniques.
Lastly, we believe time-varying
neural network is a relatively less explored domain of machine learning that has applications in both prediction and analysis of asset returns.

\bibliography{time-varying_neural_network}

\begin{thebibliography}{}

\bibitem[Abe and Nakayama, 2018]{Abe:2018}
Abe, M. and Nakayama, H. (2018).
\newblock Deep learning for forecasting stock returns in the cross-section.
\newblock In {\em arXiv}.

\bibitem[Ambachtsheer, 1974]{InfoCoeff:1974}
Ambachtsheer, K. (1974).
\newblock Profit potential in an almost efficient market.
\newblock {\em Journal of Portfolio Management}, 1(1):84.

\bibitem[Aydore et~al., 2019]{DTSSGD:2019}
Aydore, S., Zhu, T., and Foster, D.~P. (2019).
\newblock Dynamic local regret for non-convex online forecasting.
\newblock In Wallach, H., Larochelle, H., Beygelzimer, A., d'Alch\'{e} Buc, F.,
  Fox, E., and Garnett, R., editors, {\em Advances in Neural Information
  Processing Systems 32}, pages 7980--7989. Curran Associates, Inc.

\bibitem[Bergmeir et~al., 2018]{Bergmeir:2018}
Bergmeir, C., Hyndman, R., and Koo, B. (2018).
\newblock A note on the validity of cross-validation for evaluating
  autoregressive time series prediction.
\newblock {\em Computational Statistics \& Data Analysis}, 120:70--83.

\bibitem[Bossaerts and Hillion, 1999]{Bossaerts:1999}
Bossaerts, P. and Hillion, P. (1999).
\newblock Implementing statistical criteria to select return forecasting
  models: What do we learn?
\newblock {\em The Review of Financial Studies}, 12(2):405--428.

\bibitem[Cont, 2001]{Cont:01}
Cont, R. (2001).
\newblock Empirical properties of asset returns: stylized facts and statistical
  issues.
\newblock {\em Quantitative Finance}, 1:223--236.

\bibitem[Cover, 1991]{UniversalPortfolios:1991}
Cover, T.~M. (1991).
\newblock Universal portfolios.
\newblock {\em Mathematical Finance}, 1(1):1--29.

\bibitem[Cybenko, 1989]{Cybenko:1989}
Cybenko, G. (1989).
\newblock Approximation by superpositions of a sigmoidal function.
\newblock {\em Mathematics of Control, Signals and Systems}, 2(4):303--314.

\bibitem[Daniel and Moskowitz, 2016]{MomCrash:2016}
Daniel, K. and Moskowitz, T.~J. (2016).
\newblock Momentum crashes.
\newblock {\em Journal of Financial Economics}, 122(2):221--247.

\bibitem[Fabozzi et~al., 2011]{FabozziBook:2011:ch9}
Fabozzi, F.~J., Grant, J.~L., and Vardharaj, R. (2011).
\newblock {\em Common Stock Portfolio Management Strategies}, volume~1,
  chapter~9, pages 229--270.
\newblock John Wiley \& Sons, Ltd.

\bibitem[Fama and French, 1992]{FamaFrench:1992}
Fama, E.~F. and French, K.~R. (1992).
\newblock The cross-section of expected stock returns.
\newblock {\em Journal of Finance}, 47(2):427--465.

\bibitem[Gama et~al., 2014]{Gama:2014}
Gama, J.~a., \v{Z}liobait\.{e}, I., Bifet, A., Pechenizkiy, M., and Bouchachia,
  A. (2014).
\newblock A survey on concept drift adaptation.
\newblock {\em ACM Computing Surveys}, 46(4):44:1--44:37.

\bibitem[Goodfellow et~al., 2016]{Goodfellow:2016}
Goodfellow, I., Bengio, Y., and Courville, A. (2016).
\newblock {\em Deep Learning}.
\newblock MIT Press.
\newblock \url{http://www.deeplearningbook.org}.

\bibitem[Green et~al., 2017]{Green:2017}
Green, J., Hand, J. R.~M., and Zhang, X.~F. (2017).
\newblock The characteristics that provide independent information about
  average {U.S.} monthly stock returns.
\newblock {\em The Review of Financial Studies}, 30(12):4389--4436.

\bibitem[Grinold and Kahn, 1999]{GrinoldKahn:99}
Grinold, R. and Kahn, R. (1999).
\newblock {\em Active Portfolio Management: A Quantitative Approach for
  Producing Superior Returns and Controlling Risk}.
\newblock McGraw-Hill Education.

\bibitem[Gu et~al., 2020]{Gu:2020}
Gu, S., Kelly, B., and Xiu, D. (2020).
\newblock Empirical asset pricing via machine learning.
\newblock {\em The Review of Financial Studies}, 33(5):2223--2273.

\bibitem[Harvey et~al., 2016]{Harvey:2016}
Harvey, C.~R., Liu, Y., and Zhu, H. (2016).
\newblock ... and the cross-section of expected returns.
\newblock {\em The Review of Financial Studies}, 29(1):5--68.

\bibitem[Hazan et~al., 2017]{Hazan:2017}
Hazan, E., Singh, K., and Zhang, C. (2017).
\newblock Efficient regret minimization in non-convex games.
\newblock In Precup, D. and Teh, Y.~W., editors, {\em Proceedings of the 34th
  International Conference on Machine Learning, {ICML} 2017}, volume~70 of {\em
  ICML}, pages 1433--1441, Sydney, Australia. PMLR.

\bibitem[He et~al., 2016]{ResNet:2016}
He, K., Zhang, X., Ren, S., and Sun, J. (2016).
\newblock Deep residual learning for image recognition.
\newblock In Agapito, L., Berg, T., Kosecka, J., and Zelnik-Manor, L., editors,
  {\em Proceedings of the 2016 IEEE Conference on Computer Vision and Pattern
  Recognition (CVPR)}, CVPR, pages 770--778, Las Vegas, NV, USA. IEEE.

\bibitem[Hornik et~al., 1989]{Hornik:1989}
Hornik, K., Stinchcombe, M., and White, H. (1989).
\newblock Multilayer feedforward networks are universal approximators.
\newblock {\em Neural Networks}, 2(5):359–--366.

\bibitem[Horowitz et~al., 2000]{Horowitz:2000}
Horowitz, J.~L., Loughran, T., and Savin, N. (2000).
\newblock The disappearing size effect.
\newblock {\em Research in Economics}, 54(1):83--100.

\bibitem[Kingma and Ba, 2015]{ADAM:2015}
Kingma, D.~P. and Ba, J. (2015).
\newblock Adam: A method for stochastic optimization.
\newblock In Bengio, Y. and LeCun, Y., editors, {\em Proceedings of the 3rd
  International Conference on Learning Representations, {ICLR} 2015}, ICLR, San
  Diego, CA, USA.

\bibitem[LeCun et~al., 2012]{LeCun:2012}
LeCun, Y.~A., Bottou, L., Orr, G.~B., and M{\"u}ller, K.-R. (2012).
\newblock {\em Efficient BackProp}, pages 9--48.
\newblock Springer Berlin Heidelberg, Berlin, Heidelberg.

\bibitem[Lev and Srivastava, 2019]{ValueFailure:2019}
Lev, B. and Srivastava, A. (2019).
\newblock Explaining the recent failure of value investing.
\newblock In {\em SSRN}.

\bibitem[Mahsereci et~al., 2017]{Mahsereci:2017:EarlyStop}
Mahsereci, M., Balles, L., Lassner, C., and Hennig, P. (2017).
\newblock Early stopping without a validation set.
\newblock {\em CoRR}, abs/1703.09580.

\bibitem[McLean and Pontiff, 2016]{AcademicDestroy:2016}
McLean, R.~D. and Pontiff, J. (2016).
\newblock Does academic research destroy stock return predictability?
\newblock {\em The Journal of Finance}, 71(1):5--32.

\bibitem[Messmer, 2017]{Messmer:2017}
Messmer, M. (2017).
\newblock Deep learning and the cross-section of expected returns.
\newblock In {\em SSRN}.

\bibitem[Morgan and Bourlard, 1990]{Morgan:Bourland:1990}
Morgan, N. and Bourlard, H.~A. (1990).
\newblock Generalization and parameter estimation in feedforward nets: Some
  experiments.
\newblock In Touretzky, D.~S., editor, {\em Advances in Neural Information
  Processing Systems 2}, pages 630--637. Morgan-Kaufmann.

\bibitem[Pesaran and Timmermann, 1995]{PesaranTimmermann:1995}
Pesaran, M.~H. and Timmermann, A. (1995).
\newblock Predictability of stock returns: Robustness and economic
  significance.
\newblock {\em Journal of Finance}, 50:1201--1228.

\bibitem[Prechelt, 1998]{Prechelt:1998}
Prechelt, L. (1998).
\newblock Early stopping - but when?
\newblock In {\em Neural Networks: Tricks of the Trade, This Book is an
  Outgrowth of a 1996 NIPS Workshop}, pages 55--69, London, UK, UK.
  Springer-Verlag.

\bibitem[Reed, 1993]{Reed:1993}
Reed, R.~D. (1993).
\newblock Pruning algorithms-a survey.
\newblock {\em Transactions on Neural Networks}, 4(5):740--747.

\bibitem[Rosenberg et~al., 1985]{Rosenberg:1985}
Rosenberg, B., Reid, K., and Lanstein, R. (1985).
\newblock Persuasive evidence of market inefficiency.
\newblock {\em Journal of Portfolio Management}, 11(3):9--16.
\newblock Name - New York Stock Exchange; Copyright - Copyright Euromoney
  Institutional Investor PLC Spring 1985; Last updated - 2015-05-25.

\bibitem[Rossi and Inoue, 2012]{Rossi:Inoue:2012}
Rossi, B. and Inoue, A. (2012).
\newblock Out-of-sample forecast tests robust to the choice of window size.
\newblock {\em Journal of Business \& Economic Statistics}, 30(3):432--453.

\bibitem[Schroff et~al., 2015]{FaceNet:2015}
Schroff, F., Kalenichenko, D., and Philbin, J. (2015).
\newblock Facenet: A unified embedding for face recognition and clustering.
\newblock In Grauman, K., Learned-Miller, E., Torralba, A., and Zisserman, A.,
  editors, {\em Proceedings of the 2015 IEEE Conference on Computer Vision and
  Pattern Recognition (CVPR)}, CVPR, pages 815--823, Boston, MA, USA. IEEE.

\bibitem[Shalev-Shwartz, 2012]{Shalev-Shwartz:2012}
Shalev-Shwartz, S. (2012).
\newblock Online learning and online convex optimization.
\newblock {\em Foundations and Trends® in Machine Learning}, 4(2):107--194.

\bibitem[Sjöberg and Ljung, 1995]{Sjoberg:Ljung:1995}
Sjöberg, J. and Ljung, L. (1995).
\newblock Overtraining, regularization and searching for a minimum, with
  application to neural networks.
\newblock {\em International Journal of Control}, 62(6):1391--1407.

\bibitem[Sutskever et~al., 2013]{Sutskever:2013}
Sutskever, I., Martens, J., Dahl, G., and Hinton, G. (2013).
\newblock On the importance of initialization and momentum in deep learning.
\newblock In Dasgupta, S. and McAllester, D., editors, {\em Proceedings of the
  30th International Conference on Machine Learning, {ICML} 2013}, volume~28 of
  {\em ICML}, pages 1139--1147, Atlanta, Georgia, USA. PMLR.

\bibitem[Sutskever et~al., 2014]{Seq2Seq}
Sutskever, I., Vinyals, O., and Le, Q.~V. (2014).
\newblock Sequence to sequence learning with neural networks.
\newblock In Ghahramani, Z., Welling, M., Cortes, C., Lawrence, N.~D., and
  Weinberger, K.~Q., editors, {\em Advances in Neural Information Processing
  Systems 27}, pages 3104--3112. Curran Associates, Inc.

\bibitem[{U.S. Securities and Exchange Commission}, 2013]{SECMicro:2013}
{U.S. Securities and Exchange Commission} (2013).
\newblock Microcap stock: A guide for investors.
\newblock
  \url{https://www.sec.gov/reportspubs/investor-publications/investorpubsmicrocapstockhtm.html}.
\newblock Accessed: 2021-01-03.

\bibitem[Weigand, 2019]{Weigand:2019}
Weigand, A. (2019).
\newblock Machine learning in empirical asset pricing.
\newblock {\em Financial Markets and Portfolio Management}, 33:93--104.

\bibitem[Welch and Goyal, 2008]{Welch:2008}
Welch, I. and Goyal, A. (2008).
\newblock A comprehensive look at the empirical performance of equity premium
  prediction.
\newblock {\em The Review of Financial Studies}, 21(4):1455--1508.

\end{thebibliography}
\bibliographystyle{apalike}

\end{document}